\documentclass[12pt,A4]{article}

\pdfoutput=1

\usepackage[super,sort&compress,comma]{natbib}

\usepackage{times,mathptm}

\usepackage{graphicx} 

\usepackage[version=3]{mhchem} 
\usepackage{color}

\begin{document}

\noindent\LARGE{\textbf{A computational investigation of
\ce{H2} adsorption and dissociation on \ce{Au} nanoparticles
supported on \ce{TiO2} surface}}
\vspace{0.6cm}

\noindent\large{\textbf{Andrey Lyalin,\footnote{On leave from: V. A. Fock Institute of Physics, St Petersburg State University,
198504 St Petersburg, Petrodvorez, Russia.}
 and Tetsuya Taketsugu}}\vspace{0.5cm}

\noindent {{\em Division of Chemistry, Graduate School of Science,\\
Hokkaido University, Sapporo 060-0810, Japan.\\
E-mail: lyalin@mail.sci.hokudai.ac.jp}}


\vspace{0.6cm}                

\noindent \normalsize{
The specific role played by small gold nanoparticles supported on the rutile
\ce{TiO2}(110) surface
in the processes of adsorption and dissociation of \ce{H2} is discussed.
It is demonstrated that the molecular and dissociative adsorption of \ce{H2} on
\ce{Au_{\it n}} clusters containing n = 1, 2, 8 and 20 atoms
depends on cluster size, geometry structure, cluster flexibility and
the interaction with the support material.
Rutile \ce{TiO2}(110) support energetically promotes \ce{H2}
dissociation on gold clusters.
It is demonstrated that the active sites towards \ce{H2} dissociation
are located at corners and edges on the surface of the gold nanoparticle in the
vicinity of the support. The low coordinated oxygen atoms on the \ce{TiO2}(110)
surface play a crucial role for \ce{H2} dissociation.
Therefore the catalytic activity of a gold nanoparticle supported on the rutile \ce{TiO2}(110)
surface is proportional to the length of the perimeter interface between the nanoparticle
and the support.}

\vspace{0.5cm}

\section{Introduction}

Since the pioneering work of Haruta on the oxidation of carbon monoxide by small gold nanoparticles
supported by metal oxides,\cite{Haruta87} an extensive interest has been devoted
to understanding the catalytic properties of gold. Such an interest is stipulated by the
fact that gold nanoparticles are active even at room temperatures
that makes them unique catalysts for many industrial applications.\cite{Haruta03,Thompson07}

The most explored type of catalytic reactions with gold nanoparticles are
reactions of oxidation and epoxidation,
including the oxidation of carbon monoxide at mild temperatures,
alcohol oxidation, the direct synthesis of hydrogen peroxide and alkene
epoxidation;
see, e.g., refs. \citenum{Haruta87,Haruta97,Turner08,Hughes05,Tsunoyama05,Landon02,Lyalin09,Lyalin10a}
and references therein.
In spite of intensive theoretical and experimental studies the origin of catalytic activity
of gold remains highly debated.

It is commonly accepted that several factors can influence the catalytic activity of gold.
The most important factor is the size effect. It has been shown that the unique properties
of gold in oxidation reactions emerge when the size of catalytic particles decreases
down to 1-5 nm or even less; while larger sized particles and the bulk form of gold
are catalytically inactive.\cite{Haruta87,Haruta97,Tsunoyama05,Turner08}
It has been shown that small gold clusters consisting of a few atoms can also possess
extraordinarily high catalytic activity.\cite{Herzing08,Maria08}
On the one hand, the size effects in gold nanocatalysis are
determined by quantum effects, resulting from the spatial
confinement of the valence electrons in the cluster;\cite{Nanocat07}
on the other hand, in such clusters a dominant fraction of atoms are under-coordinated
(in comparison with the bulk), hence they exhibit an enhanced chemical
reactivity.\cite{Nanocat07,Hvolbaek07,Lang09}

Interaction with the support material is another important factor that considerably
influences the chemical reactivity of the gold
nanoparticles.\cite{Haruta97,Landman07,Harding09,Rodriguez10}
Most experimental studies on the catalytic properties of gold have been
performed for gold nanoparticles supported on the surfaces of metal oxides; see, \emph{e.g.},
refs. \citenum{Haruta97,Coquet08} and references therein.
It has been demonstrated that the catalytic activity of gold
nanoparticles depends on the presence of defects in the support material (e.g. F-center defects),
charge transfers from the support, the chemical state of the surface, doping,
{moisture}, special additives, and the presence of adsorbates on the cluster surface,
including the reactant molecule itself; see, e.g.,
refs. \citenum{Haruta97,Sanchez99,Date04,Landman07,Matthey07,Coquet08,Joshi06,Lyalin10,Lyalin10a}
and references therein. 
However, a work by Turner \emph{et al.} presents strong experimental evidence that
small gold entities ($\sim$ 1.4 nm) derived  from \ce{Au_{55}} clusters and
deposited on an inert support can also be efficient and robust
catalysts for oxidation reactions.\cite{Turner08}
Hence, the interaction with the support is important but not the determining factor in catalytic
activity of gold; therefore even free clusters can be effective catalysts.

Heterogeneously catalyzed hydrogenation is another type of reactions
where gold nanoparticles have shown their great potential as catalysts.
It has been demonstrated experimentally that gold nanoparticles supported on metal oxides such as
\ce{SiO2}, \ce{Al2O3}, \ce{TiO2}, \ce{ZnO}, \ce{ZrO2}, and \ce{Fe2O3}
are effective catalysts for selective hydrogenation of several classes of organic molecules,
including $\alpha$, $\beta$-unsaturated aldehydes, unsaturated ketones, and unsaturated
hydrocarbons.\cite{Jia00,Choudhary03,Bailie99,Schimpf02,Okumura02,Mohr03,Zanella04,Claus05,Zhu10,Zhu10a}
Moreover, supported gold nanoparticles are very selective for the direct
formation of hydrogen peroxide from \ce{H2}/\ce{O2} mixtures.\cite{Landon02}
Mechanisms of such catalytic reactions are largely not understood.
Experimental results demonstrate that molecular hydrogen does not bind
to the clean gold extended surface, and molecular or dissociative hydrogen adsorption
occurs on the supported gold nanoparticles.\cite{Claus05,Bus05}
It has been suggested that hydrogen reacts and dissociates on low coordinated
gold atoms in the corner or at edge positions on the surface of gold nanoparticles.\cite{Bus05,Mohr03}
It has been also shown that the shape of the gold particles plays a role on
their catalytic activity.\cite{Mohr03a}
However, in a recent work by Haruta {\it et al} it has been
found that the catalytic activity of the gold nanoparticles supported on the rutile \ce{TiO2}(110)
surface depends on the number of gold atoms located at the perimeter interface of
the supported gold nanoparticles, {irrespective of} the cluster size.\cite{Fujitani09}
It was also supposed that active sites for \ce{H2} dissociation may be formed
by a combination of gold atoms and oxygen atoms from \ce{TiO2} at nanoparticle--surface
interface.\cite{Fujitani09}

Surprisingly, very little attention has been paid to theoretical investigations of the
hydrogenation reactions on gold nanoparticles. 
Interaction of molecular hydrogen with gold atoms and small gold clusters
has been theoretically studied in refs.
\citenum{Andrews04,Andrews04a,Zanchet10,Varganov04,Okumura05,Barrio06,Ghebriel07,Joshi07}.
It has been demonstrated, that the flexibility of the cluster structure plays a key role
in the bonding and dissociation of \ce{H2} on \ce{Au_{14}} and \ce{Au_{29}}.\cite{Barrio06}
Furthermore, authors in ref. \citenum{Barrio06} have concluded that dissociation
of \ce{H2} is only possible when a cooperation between four active
low coordinated Au atoms on the cluster surface is allowed.
However, in ref. \citenum{Corma07} the compelling evidence has been presented that \ce{H2} adsorbs
and dissociates with small activation barriers on low coordinated Au atoms
situated in corner positions of different \ce{Au_{25}} nanoparticles without further conditions.
Thus, there is no need for cooperation between several active Au atoms to dissociate \ce{H2}
and dissociation can occur spontaneously on a single low coordinated Au atom, when
coordination number is four.\cite{Corma07}

There are relatively few theoretical studies on the catalytic hydrogenation reactions on gold clusters. 
These include works on formation of hydrogen peroxide from \ce{H2} and \ce{O2} over  
small gold clusters;\cite{Wells04,Kacprzak09} acrolein hydrogenation on \ce{Au20} 
nanoparticle;\cite{Li10} and combined experimental and theoretical investigation 
of the unusual catalytic
properties of gold nanoparticles in the selective hydrogenation of 1,3-butadiene toward
cis-2-butene.\cite{Yang10}


Despite of the clear experimental evidence {indicating a strong influence} of the support materials
on the processes of adsorption and dissociation of molecular hydrogen and related hydrogenation
processes on the supported gold nanoparticles,\cite{Claus05,Fujitani09} there are no
theoretical studies concerning the role of the support
with an exception of recent works by Boronat et al.\cite{Boronat09}
and Florez et. al.\cite{Florez10}
In ref. \citenum{Boronat09} authors performed an elegant theoretical study on
elucidation of active sites for \ce{H2} adsorption and activation
on \ce{Au_{13}} cluster supported on the anatase \ce{TiO2} (001) surface.
It has been shown that Au atoms that are active for \ce{H2} dissociation must have a net charge
close to zero, be located at corner or edge low coordinated positions, and
not be directly bonded to the support.\cite{Boronat09} The \ce{Au13} isomers consisting
of two layers of gold atoms have the largest number of potentially active sites where
\ce{H2} can be adsorbed and activated. These active atoms are located on the top layer
of the \ce{Au_{13}} cluster that does not have contact with the support.\cite{Boronat09}
It has been demonstrated that the presence of O vacancy defects on the anatase
\ce{TiO2} (001) surface can stabilize the most active two-layer \ce{Au_{13}}
particles.\cite{Boronat09} In ref. \citenum{Florez10} it was shown that small
two-dimensional Au nanoparticles supported on TiC(001) can dissociate \ce{H2}
in a more efficient way than when supported on oxides. The active sites
for \ce{H2} dissociation on Au/TiC(001) are located at the particle edge and in direct contact
with the underlying substrate.\cite{Florez10}


In the present paper we report the results of a systematic theoretical study
of the adsorption and dissociation of \ce{H2} on free and \ce{TiO2}(110) supported
\ce{Au_{\it n}} clusters containing n = 1, 2, 8 and 20 atoms.
In our work we want to clarify how hydrogen molecule adsorbs and dissociates on
the gold nanoparticles. What factors are the most important for \ce{H2}
dissociation on the surface of gold? { Does the} dimensionality of the cluster
play a role in \ce{H2} adsorption and dissociation?
What is the role of the support? Is it possible to promote dissociation of
\ce{H2}?
Whether or not the charge transfer between the adsorbed \ce{H2} and the gold cluster
plays any role in \ce{H2} catalytic activation and dissociation?
Why the catalytic activity of gold nanoparticles
supported on the rutile \ce{TiO2}(110) surface
depends on the number of gold atoms located at the perimeter interface,
irrespective to the particle size in the case of \ce{H2} dissociation,
but it depends on cluster size in the case of \ce{O2} dissociation?
Finally, what are the important directions for further
theoretical investigation of the considered processes?
Answering these questions is the central aim of the present work.


\section{Computational Details}

The calculations are carried out using density-functional theory (DFT)
with the gradient-corrected exchange-correlation functional of
Perdew, Burke and Ernzerhof (PBE).\cite{Perdew96}
The atom-centered, strictly confined, numerical basis functions\cite{Artacho99,Junquera01}
are used to treat the valence electrons of H, Au, Ti and O atoms
with the reference configurations
$1s^1 2p^0 3d^0    4f^{0}$ for H,
$6s^1 6p^0 5d^{10} 5f^0$ for Au,
$4s^2 4p^0 3d^{2}  4f^{0}$ for Ti,  and
$2s^2 2p^4 3d^{0}  4f^{0}$ for O.
Double-$\zeta$ plus polarization function (DZP) basis sets are used for
Au, Ti and O, and  triple-$\zeta$ plus polarization function (TZP) for H.
The remaining core electrons are represented by the Troullier-Martins norm-conserving
pseudopotentials\cite{Troullier91} in the Kleinman-Bylander factorised form.\cite{Kleinman82}
Relativistic effects are taken into account for Au.

Calculations have been carried out with the use of the
SIESTA code.\cite{Sanchez-Portal97,Soler02,Sanchez-Portal04}
Periodic boundary conditions are used for all systems, including free molecules and clusters.
In the latter case the size of a supercell was chosen to be large enough to make
intermolecular interactions negligible.

Within the SIESTA approach, the basis functions and the electron density are projected
onto a uniform real-space grid. The mesh size of the grid is controlled by an energy cutoff,
which defines the wavelength of the shortest plane wave that can be represented on the grid.
In the present work the energy cutoff of 200 Ry is chosen to guarantee convergence of the
total energies and forces. The self-consistency of the density matrix is achieved with a
tolerance of $10^{-4}$. For geometry optimization the conjugate-gradient
approach was used with a threshold of 0.02 eV \AA$^{-1}$.

The range of the basis pseudoatomic orbitals (PAO) is limited by an energy shift
parameter which defines the confinement of the basis functions.
It has been shown that systematic convergence to physical quantities can be obtained
by varying the energy shift.\cite{Artacho99,Junquera01}
A common energy shift of 50 meV is applied for Au, Ti and O on the basis of careful
comparison of the obtained theoretical results with the available reference data
for the free \ce{Au2} dimer, small gold clusters, bulk Au and \ce{TiO2}.
Thus the calculated values of the dissociation energy and bond length in \ce{Au2}
(2.29 eV, 2.572 \AA)
are in an excellent agreement with those of earlier experimental studies,
({2.31} eV,  2.472 \AA).\cite{Herzberg79}
The optimized lattice constant of the face-centered cubic (fcc) structure of bulk gold is
4.208  \AA, which is 3\% larger than the experimental value of
4.079 \AA.\cite{Morosin68} This is a general feature of PBE functional to
overestimate the lattice constants for various types of solids.\cite{Haas09}
In addition we tested the ability of our approach to reproduce the optimized
structures and isomer sequences of small free gold clusters consisting of up to 10 atoms.
The obtained structures are in a good agreement with those reported in our
recent work ref. \citenum{Lyalin10} and previous theoretical studies; see, e.g.,
refs. \citenum{Ding04,Fernandez04,Walker05,Xiao06,Hakkinen08}.

Basis set for hydrogen was optimized  with the use of the
Nelder-Mead simplex method\cite{Nelder65} according to the procedure
described in ref. \citenum{Junquera01}.
The dissociation energy, $D_e$, and bond length in \ce{H2} { (4.55 eV, 0.750 \AA) }
are in a good agreement with experimental data 
(4.74 eV, 0.741 \AA).\cite{Herzberg79}
We also tested the gold -- hydrogen interaction  optimizing \ce{AuH} dimer.
The calculated dissociation energy and bond length in \ce{AuH} (3.09 eV,
1.547 \AA) are in a good agreement with experimental data
({ 3.36} eV, 1.524 \AA).\cite{Herzberg79}

The rutile structure belongs to the $P4_2/mnm$ tetragonal space group.
The primitive unit cell contains two \ce{TiO2} units.
The unit cell parameters determined by neutron diffraction
are found to be a=b=4.587 \AA, c=2.954 \AA\ at 15 K.\cite{Burdett87,Muscat02}
In the present work the rutile lattice was optimized
using the Monkhorst-Pack\cite{Monkhorst-Pack} 6$\times$6$\times$9 k-point mesh 
for Brillouin zone sampling.
The calculated lattice parameters a=4.594 \AA\ and c=2.959 \AA\
are in excellent agreement with the experimental values.

The optimized lattice of the bulk rutile was used to construct
slabs for the \ce{TiO2}(110) surface. The six-layer
slab containing four units of \ce{TiO2} represents the element of the
(110) rutile face with the surface area of $\sqrt 2$a$\times$c.
In the present work we study adsorption of \ce{Au_{\it n}} clusters
containing n = 1, 2, 8 and 20 atoms on the rutile (110) surface.
In the case of Au and \ce{Au2} the \ce{TiO2}(110) surface is modeled by
the p(2$\times$5) slab (40 units of \ce{TiO2} per slab), while for
\ce{Au8} and \ce{Au20} we use p(3$\times$6) (72 units of \ce{TiO2})
and p(4$\times$9) (144 units of \ce{TiO2}) slabs, respectively.
The periodically replicated slabs are separated by the vacuum
region of 25 \AA\ in the (110) direction.
It has been shown that the surface energy, the energy gap and interlayer distances
in thin films of rutile oscillate with decreasing amplitude as the number of
layers increases.\cite{Perron07,Bredow04,Ramamoorthy94}
The choice of the six-layer slab provides good convergence of the obtained
results at moderate computational cost.
Similar slab model has been used in ref. \citenum{Liu06} to study an \ce{O2}
supply pathways in \ce{CO} oxidation on Au/\ce{TiO2}(110).

\begin{figure}[htbp]
\begin{center}
\includegraphics[scale=0.4,clip,bb=14 435 530 705]{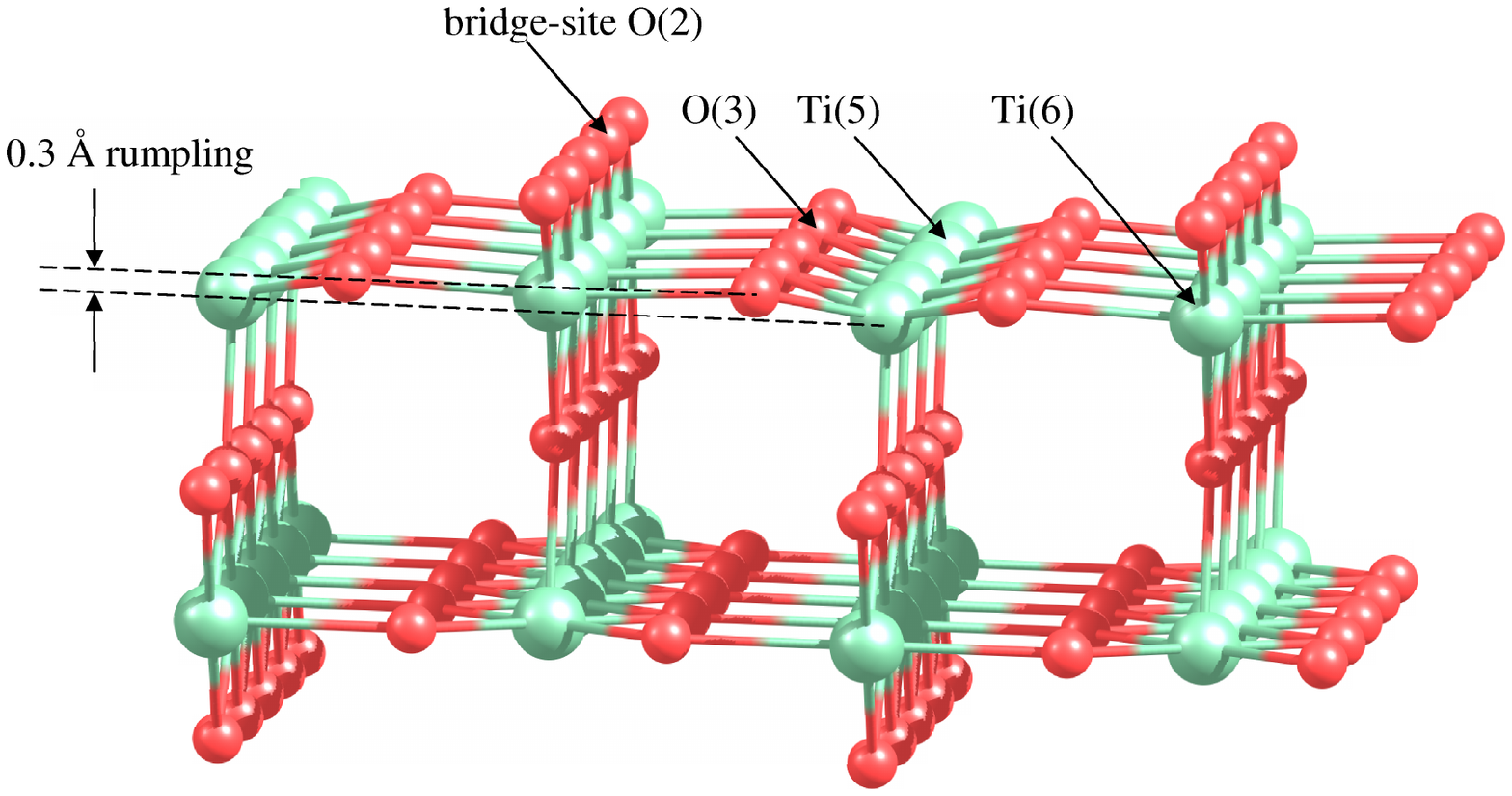}
\end{center}
\caption{Side view of the \ce{TiO2}(110) p(2$\times$5) slab. The bridge-site twofold coordinated O(2)
atom, threefold coordinated O(3) atom, five- and sixfold coordinated Ti(5) and Ti(6) atoms
are marked, respectively. Relaxation of the \ce{TiO2}(110) surface
results in a surface rumpling.}
\label{fig:surface_structure}
\end{figure}

Fig. \ref{fig:surface_structure} presents the side view of the \ce{TiO2}(110) p(2$\times$5) slab.
The \ce{TiO2}(110) surface contains atoms with different coordination.
The fivefold coordinated Ti atoms are undercoordinated
relatively to the bulk sixfold coordinated Ti.
The O atoms raised above the plane of Ti atoms are
twofold coordinated. These atoms are called bridge-site O.
The oxygen atoms { lying } in the plane of Ti atoms are threefold coordinated.
In all calculations the bottom two layers in the slabs are fixed,
and all other atoms are fully relaxed. Relaxation of the \ce{TiO2}(110) surface
results in a surface rumpling, as it is shown in Fig. \ref{fig:surface_structure}.
The undercoordinated Ti and O atoms are moving inward by 0.10 \AA,
decreasing the distance with subsurface atoms.
The sixfold coordinated Ti atoms and threefold coordinated O atoms
are moving outward by 0.1 \AA\ and 0.2 \AA, respectively. The similar
puckered structure of the rutile (110) surface
has been reported in refs. \citenum{Ramamoorthy94,Perron07}.

\section{Results}

\subsection{Hydrogen adsorption on free gold clusters}

The molecular and dissociative adsorption of \ce{H2}
on the free gold clusters of different sizes, chains of gold atoms, and
extended gold surfaces have been the subject of a number of theoretical
studies.\cite{Varganov04,Wang06,Barrio06,Kang09a,Zanchet09,Zanchet10,Corma07}
It was demonstrated that \ce{H2} effectively adsorbs and dissociates at the
low coordinated Au atoms regardless if they are in gold clusters
or at extended line defects, such as monoatomic rows on
Au surface.\cite{Corma07,Barrio06}  However, in the case of small
Au clusters, coordination alone cannot explain all the features of the reactivity
of gold clusters for \ce{H2} dissociation.\cite{Zanchet09}  Several
major factors are determining the catalytic properties of free clusters:
the structure associated with individual atoms (i.e. coordination), the binding structure, and
the border effects.\cite{Zanchet09} {It was demonstrated that
the planar gold structures are not reactive, when the \ce{H2} approaches the cluster
from the top.} However, dissociation of \ce{H2}
can be promoted when planar clusters are folded or the \ce{H2} approaches
clusters in their plane.\cite{Zanchet09} The flexibility can play an important
role in \ce{H2} dissociation, because of the large atomic displacements
in the cluster.\cite{Barrio06}  Finally, the size, the structure and the charge state
of gold clusters influence on the molecular and dissociative adsorption
of \ce{H2}.\cite{Kang09a}

In the present work we consider adsorption of \ce{H2} on gold clusters
\ce{Au_{\it n}} consisting of $n=$ 1, 2, 8, and 20 atoms in order to elucidate the size
and the geometry effects on hydrogen adsorption.
The structural properties of free gold clusters have been intensively studied,
see, e.g., refs. \citenum{Pyykko08,Coquet08,Hakkinen08} for a review.
DFT calculations demonstrate that \ce{Au_{\it n}} clusters favor planar,
two-dimensional (2D) structures up to cluster size of 11-13 atoms,
while \ce{Au_{\it n}} of larger sizes are three-dimensional
(3D).\cite{Hakkinen03,Xiao06,Ass09}  The high stability of planar
structures for small gold clusters has been explained by relativistic
effects. The relativistic effects make the $6s$ and $5d$ atomic orbitals closer in energy,
resulting in a $sd$ hybridization, which in turn stabilizes planar
configurations.\cite{Hakkinen02}
The exact value of the critical size for  2D $\rightarrow$ 3D transition in
neutral \ce{Au_{\it n}} is not yet clarified.
{\it Ab initio} MP2 and CCSD(T) calculations performed
in ref. \citenum{Olson05} demonstrate that \ce{Au8} possesses 3D structure,
while the similar CCSD(T) calculations performed with the larger basis set
favors 2D structure for \ce{Au8}.\cite{Han06} Recent CCSD(T) calculations
accounting for the basis-set superposition error
(BSSE) corrections are in favor of 3D structure for \ce{Au_{10}}\cite{Choi09}
in accord with DFT predictions.
Competition of 2D and 3D structures
of the small gold clusters is a very interesting
feature that can play an important role in cluster reactivity.
The 2D structures contain a large number of low coordinated atoms,
hence they might possess higher catalytic activity
in comparison with the 3D clusters.
{Moreover, it has been shown that if the shape of neutral gold 
clusters is modified, it is possible to change their electron donor-acceptor 
capacities -- the 2D gold clusters are better electron acceptors
than 3D ones.\cite{Martinez10} 
Since the electron transfer between the cluster and the 
adsorbate is an important driving force for the reactivity, 
the shape of the cluster can directly affect cluster's 
catalytic properties.\cite{Martinez10}}

In our previous works we have optimized the most stable and isomer geometries
of small gold clusters consisting up to 10 atoms within DFT B3PW91/LANL2DZ
approach.\cite{Lyalin09,Lyalin10} In the present work we have re-optimized the earlier
obtained geometries of \ce{Au_{\it n}} using the PBE/DZP SIESTA method.
The obtained structures are in an excellent agreement with those reported
in previous theoretical studies; see, e.g.,
refs. \citenum{Ding04,Fernandez04,Walker05,Xiao06,Hakkinen08,Lyalin10}.

\begin{figure}[htbp]
\begin{center}
\includegraphics[scale=0.5,clip,bb=65 100 565 740]{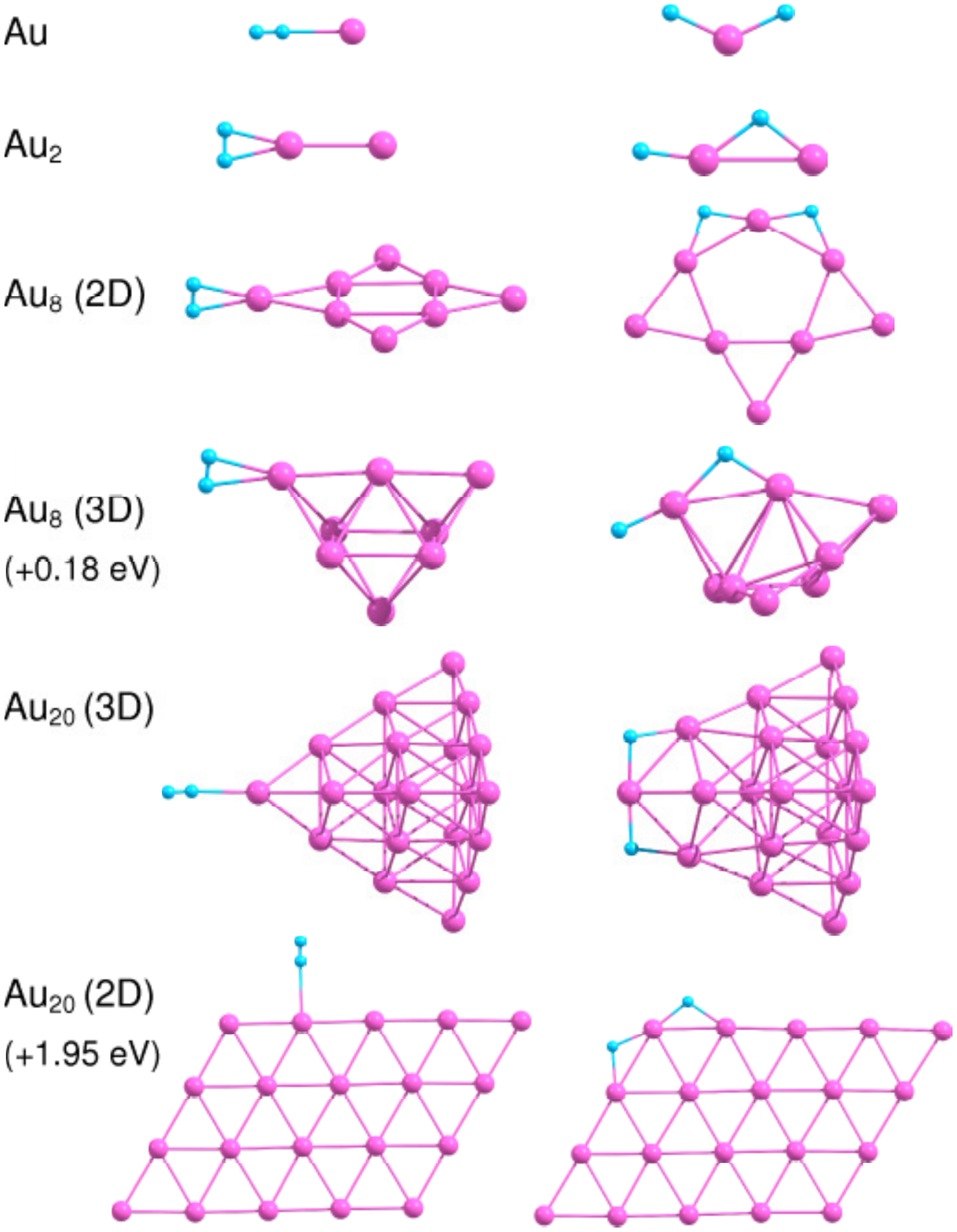}
\end{center}
\caption{Optimized geometries of \ce{H2 - Au_{\it n}} clusters in the case of molecular
(left) and dissociative (right) adsorption of \ce{H2}.
The numbers in the parentheses correspond to the
relative energy for the higher isomeric free \ce{Au8} (3D) and \ce{Au_{20}} (2D)
clusters with respect to the corresponding \ce{Au8} (2D) and \ce{Au_{20}} (3D)
lowest energy structures.}
\label{fig:adsorption_free}
\end{figure}

In Fig. \ref{fig:adsorption_free} we present optimized geometries of
\ce{H2 - Au_{\it n}} systems in the case of molecular and dissociative
adsorption of \ce{H2} on free \ce{Au_{\it n}} clusters.
In order to obtain the most stable configuration of \ce{H2}
adsorbed on \ce{Au_{\it n}}, we have created a large number of starting
geometries by adding \ce{H2} molecule in different positions (up to 30) 
on the surface of the considered clusters. The starting structures have been
optimized further without any geometry constraints.
We have successfully used a similar approach to find the optimized geometries
of \ce{O2} and \ce{C2H4} molecules adsorbed on small neutral, anionic and cationic
gold clusters.\cite{Lyalin09,Lyalin09a,Lyalin10,Lyalin10a}
In addition to the most stable 2D structure of \ce{Au8} and 3D structure of \ce{Au_{20}}
we consider the closest in energy 3D isomer of \ce{Au8} and 2D isomer of \ce{Au_{20}}
in order to understand how dimensionality and structural properties affect
the cluster reactivity. The calculated relative energies for the higher isomeric
\ce{Au8} (3D) and \ce{Au_{20}} (2D) structures with respect to the corresponding
lowest energy structures are 0.18 eV and 1.95 eV, respectively.
The binding energies for molecular, $E_{b}^{mol}$, and dissociative,
$E_{b}^{dis}$, adsorption of \ce{H2} on free \ce{Au_{\it n}} as well as
the \ce{H-H} bond length $r_{\rm H-H}^{mol}$ in \ce{H2} adsorbed molecularly
are summarized in Table \ref{tbl:binding_free}.

\begin{table}[h]
\small
\centering
  \caption{~Binding energies for molecular, $E_{b}^{mol}$,
and dissociative, $E_{b}^{dis}$, adsorption of \ce{H2} on free \ce{Au_{\it n}} clusters
and the \ce{H-H} bond length, $r_{\rm H-H}^{mol}$,
in the case of molecular adsorption of \ce{H2}.}
  \label{tbl:binding_free}
  \begin{tabular}{l c c c}
    \hline
    Cluster           & $r_{\rm H-H}^{mol}$ (\AA) & $E_{b}^{mol}$ (eV)  & $E_{b}^{dis}$ (eV)\\
    \hline
    \ce{Au}             &  0.780 & 0.13     & 0.04 \\
    \ce{Au2}            &  0.845 & 0.59     & 0.59 \\
    \ce{Au8} (2D)       &  0.806 & 0.26     & 0.68 \\
    \ce{Au8} (3D)       &  0.788 & 0.13     & 0.52   \\
    \ce{Au_{20}} (3D)   &  0.768 & 0.09     & 0.14 \\
    \ce{Au_{20}} (2D)   &  0.762 & 0.09     & 0.72  \\
    \hline
  \end{tabular}
\end{table}

The binding energy of \ce{H2} adsorbed on a free \ce{Au_{\it n}} cluster is defined as
\begin{equation}
E_{b}^{mol,dis} =  E_{tot}(\ce{Au_{\it n}}) + E_{tot}(\ce{H2}) -
               E_{tot}^{mol,dis}(\ce{H2 - Au_{\it n}}),
\end{equation}
\noindent where $E_{tot}^{mol,dis}(\ce{H2 - Au_{\it n}})$ denotes the total energy of
the compound system, \ce{H2 - Au_{\it n}},
while $E_{tot}(\ce{Au_{\it n}})$  and $E_{tot}(\ce{H2})$
are the total energies of the non-interacting fragments \ce{Au_{\it n}} and \ce{H2},
respectively.

Results of our calculations demonstrate, that \ce{H2} binds weakly to Au as a molecule,
with $E_{b}^{mol}$=0.13 eV.  Dissociation of \ce{H2} on \ce{Au} is not favorable energetically.
The \ce{H-H} bond length in \ce{H2} adsorbed on \ce{Au} is slightly enlarged
in comparison with the free \ce{H2}.
Adsorption of \ce{H2} on \ce{Au2} is relatively strong, with the binding energy
equal to 0.59 eV both for the molecular and the dissociative adsorption.
Table \ref{tbl:binding_free} demonstrates that the \ce{H-H} bond length increases to
0.845 \AA\ in \ce{H2-Au2}, indicating that \ce{H2} is highly activated.

With the further increase in cluster size the binding energy calculated for molecular
adsorption of \ce{H2}  decreases to 0.26 eV (0.13eV) for  \ce{Au8} (2D)  (\ce{Au8} (3D)) and
0.09 eV both for 3D and 2D isomers of \ce{Au_{20}}.
On the other hand, the dissociative adsorption of \ce{H2} becomes energetically favorable
both for \ce{Au8} and \ce{Au_{20}}.
The binding energy of \ce{H2} on free gold clusters strongly depends
on the cluster's isomer state. Thus, dissociative adsorption of \ce{H2}
on 2D structures of \ce{Au8} and \ce{Au_{20}} are energetically favorable compared
with the corresponding 3D structures.
Therefore, dissociation of \ce{H2} on small gold clusters
can be promoted if we can find a way to enhance the stability of 2D structures
for n $\ge$ 13.  That should be possible for supported clusters, where
an attractive interaction with the surface stabilizes oblate
configurations.\cite{Semenikhina08} Thus, varying the interaction with the
support one can change the cluster shape, and hence it can be possible to tune
its reactivity.
Another factors that influence \ce{H2} dissociation are coordination of
Au atoms interacting with hydrogen and flexibility of cluster structure.
It is seen from Fig. \ref{fig:adsorption_free}
that \ce{H2} dissociates at the low coordinated corner Au atom with
formation of the slightly bended \ce{H-Au-H} bond.
It is worth to note, that the  \ce{Au-Au} bond length in \ce{Au2} does not change
noticeably upon molecular adsorption of \ce{H2}; but increases up to 2.753 \AA\ for
dissociative adsorption.  In the case of \ce{Au8} and \ce{Au_{20}} dissociation of
\ce{H2} on the cluster surface is accompanied by the structural (at least local)
transformations. Therefore we can confirm conclusion made in ref. \citenum{Barrio06}
that the structural flexibility of the gold clusters
might be an important factor for \ce{H2} dissociation.

\subsection{Hydrogen adsorption on \ce{Au_{\it n}}/\ce{TiO2}(110)}

How does the interaction with the support influence the reactivity of gold clusters?
Can a supported nanoparticle exhibit novel 
properties that are not found for free particles?
Recent experiments performed by Haruta {\it et al} clearly demonstrate that
the perimeter interface of the gold nanoparticles supported on
the rutile \ce{TiO2}(110) surface plays a crucial role in the process of
hydrogen dissociation.\cite{Fujitani09} It was suggested that
the active sites for \ce{H2} dissociation might be formed
by a combination of gold atoms and oxygen atoms from \ce{TiO2} at the
nanoparticle--surface interface.\cite{Fujitani09}

In the present work we perform a systematic theoretical study of the structure and energetics
of \ce{H2} adsorbed on \ce{Au_{1}}, \ce{Au_{2}}, \ce{Au_{8}} (2D), \ce{Au_{8}} (3D),
\ce{Au_{20}} (3D) and \ce{Au_{20}} (2D) clusters supported
on the rutile \ce{TiO2}(110) surface. The analysis of the energetics of
\ce{H2} adsorption on the supported gold clusters provides deep insights into the
nature of bonding and dissociation of the \ce{H2} molecule and reveals the details
of the reaction mechanism.

As discussed above the dissociation of \ce{H2} on free gold clusters is governed by
the interplay of structural factors, coordination of the adsorption center and
cluster flexibility. The complexity of the changes in structural and electronic
features of gold nanoparticles becomes more diverse when one deposit nanoparticle
on the support.

In order to obtain the most stable geometries of \ce{Au_{\it n}}
supported on the rutile \ce{TiO2}(110) surface we have created
a large number of starting configurations by adding \ce{Au_{\it n}}
isomers with different orientation in respect to the surface. These structures
have been fully optimized on the rutile surface, with accounting for the
relaxation of all the gold atoms as well as the top four layers of the six
layer slab of the rutile. The bottom two layers in the slab were fixed.
Thus, we have taken into account deformations of the cluster structure
due to the interaction with the support as well as structural
relaxations on the rutile \ce{TiO2}(110) surface due to its interaction with
the supported cluster.

\begin{figure}[htbp]
\begin{center}
\vspace*{1cm}
\includegraphics[scale=0.6,clip,bb=10 10 565 340]{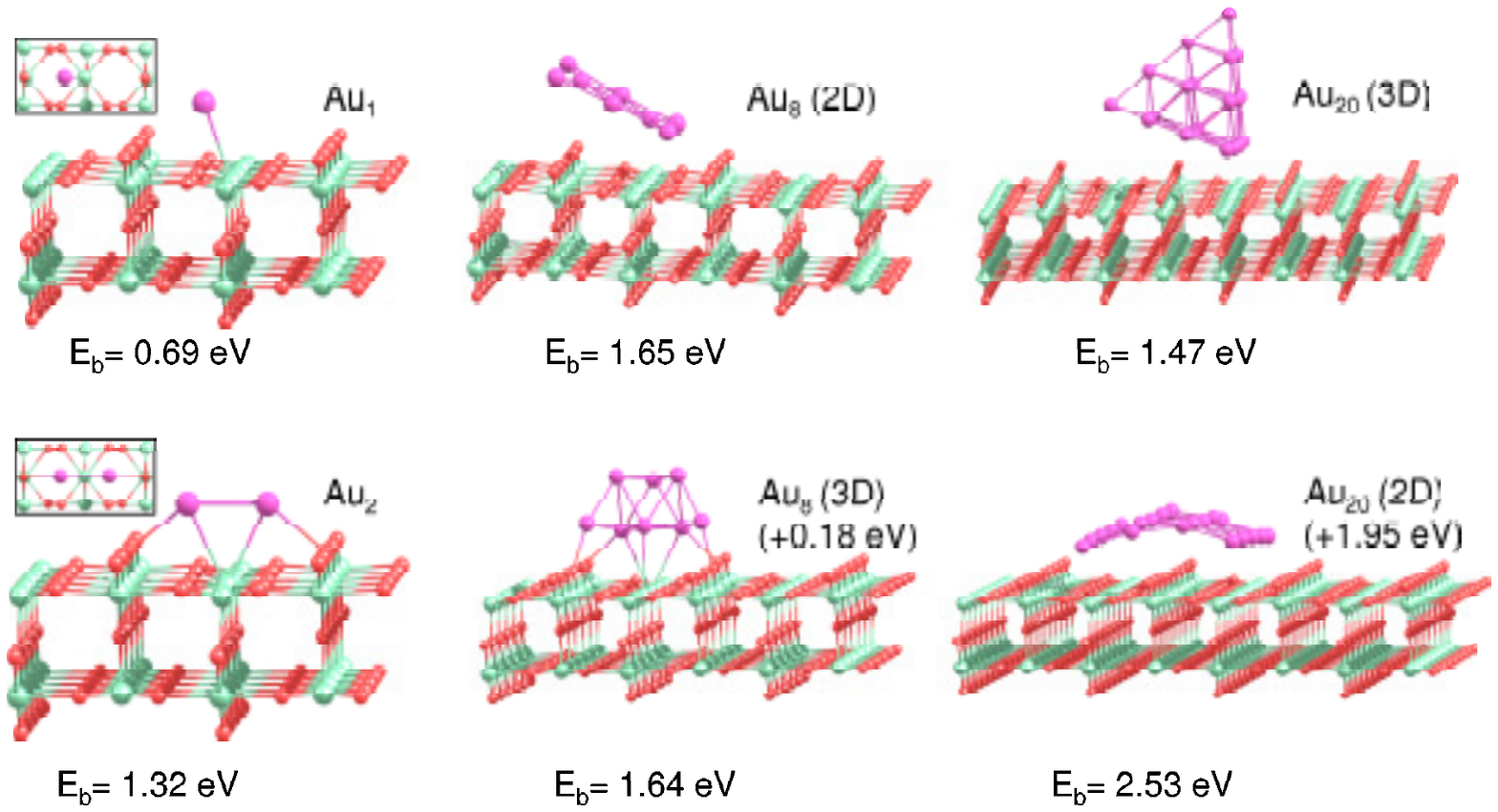}
\end{center}
\caption{The most stable geometries of \ce{Au_{\it n}} clusters ($n=$ 1, 2, 8, and 20)
optimized on the rutile \ce{TiO2}(110) surface.
Top views of \ce{Au} and \ce{Au2} on \ce{TiO2}(110) are shown in the inserts.
The binding energies of \ce{Au_{\it n}} on \ce{TiO2}(110)
are marked below the corresponding structures.
Numbers in parentheses correspond to the
relative energy for the higher isomeric free \ce{Au8} (3D) and \ce{Au_{20}} (2D)
clusters with respect to the corresponding free \ce{Au8} (2D) and \ce{Au_{20}} (3D)
lowest energy structures.}
\label{fig:Aun_TiO2}
\end{figure}

Fig. \ref{fig:Aun_TiO2} presents the most stable geometries
of \ce{Au_{\it n}} clusters ($n=$ 1, 2, 8, and 20)
optimized on the rutile \ce{TiO2}(110) surface.
We found that gold tends to maximize interaction with the O(3) and Ti(5)
atoms on \ce{TiO2}(110) surface. Thus, \ce{Au} atom tends to occupy a position above
the row of O(3) surface atoms, simultaneously bridging two O(3) atoms in the direction
parallel to the O(3) row, as well as Ti(5) and O(2) atoms, in the direction perpendicular to the
O(3) row. In the case of \ce{Au2}, both of Au atoms occupy positions above of the rows of
O(3) atoms, symmetrically in respect to the row of Ti(5) atoms. The \ce{Au-Au} bond length of
2.57 \AA\ in \ce{Au2} matches the distance between the two rows of O(3) atoms (2.53 \AA\ ).
The geometry structures of two- and three dimensional isomers of \ce{Au8} on the
\ce{TiO2}(110) surface are slightly deformed if compared with the free gold clusters.
The \ce{Au8} (2D) cluster maximizes the interaction between the cluster edge and the
row of O(3) atoms on the \ce{TiO2}(110) surface.
The angle between the cluster's plane and the rutile surface is 26$^\circ$.
The supported \ce{Au8} (3D) isomer prefers to maximize its
interaction with the rutile \ce{TiO2}(110) surface,
orienting the 5-atom face parallel to the support,
as shown in Fig. \ref{fig:Aun_TiO2}.  The binding energies of \ce{Au8} (2D) and \ce{Au8} (3D)
isomers on  \ce{TiO2}(110)  are 1.65 eV and 1.64 eV, respectively.
A tetrahedral \ce{Au_{20}} (3D) cluster binds to the surface, maximizing interaction
between the cluster edge and rows of O(3) and Ti(5) surface atoms; while  \ce{Au_{20}} (2D)
isomer binds by two its edges to two parallel rows of Ti(5) surface atoms, slightly
bending its planar structure, to avoid interaction with the row of O(2)
bridge atoms in the middle. The planar \ce{Au_{20}} structure binds to the surface 
much more strongly 
than the 3D structure, because it has a wider contact area.
Although \ce{Au_{20}} (3D) structure is still energetically
favorable when adsorbed on the \ce{TiO2}(110) surface in comparison with the 2D structure,
the energy difference between two configurations decreases from 1.95 eV for free clusters to
0.89 eV for supported clusters, respectively.
If we could further increase the interaction with the support,
we would make 2D structure energetically favorable.
Such an effect has been reported for \ce{Au_{20}} clusters deposited
on thin MgO(100) films supported on Mo(100) substrate.\cite{Ricci06}
By changing the thickness of the metal-oxide film, one can tune the interaction of the
cluster with the support, and thus increase its wetting propensity and
induce a dimensionality crossover from 3D cluster structures on MgO(100)
to the energetically favored 2D geometries on the metal-supported films.\cite{Ricci06}

\begin{figure}[htbp]
\begin{center}                     
\includegraphics[scale=0.6,clip,bb=60 540 510 670]{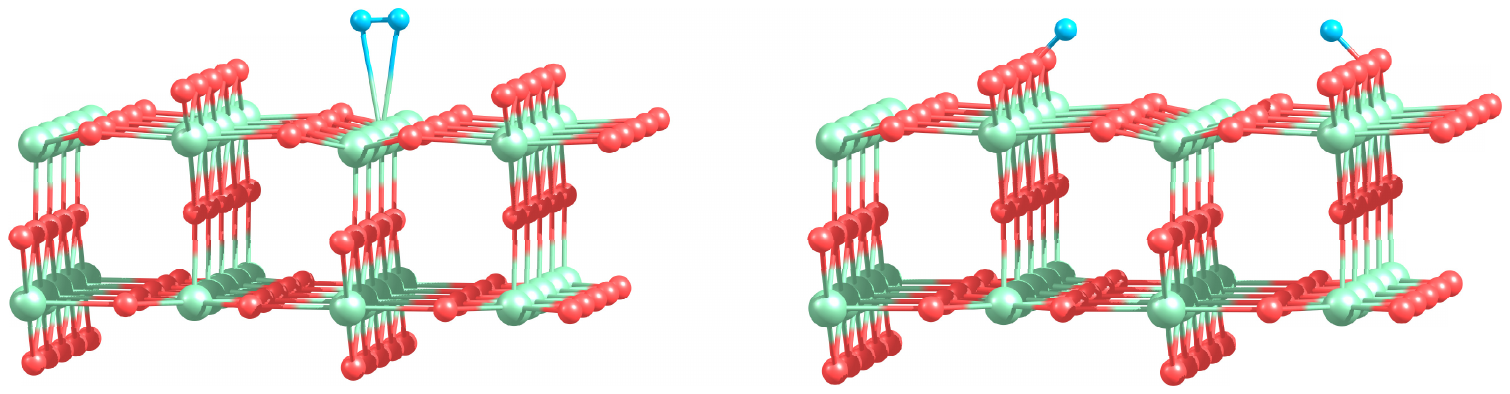}
\end{center}
\caption{Optimized geometries for molecular (left) and dissociative (right)
adsorption of \ce{H2} on the pure rutile \ce{TiO2}(110) surface.}
\label{fig:H2_TiO2}
\end{figure}

Fig. \ref{fig:H2_TiO2} presents the most stable geometries calculated for
the molecular and dissociative adsorption
of \ce{H2} on the pure rutile \ce{TiO2}(110) surface.
Our calculations demonstrate that the hydrogen molecule adsorbs on the pure
rutile \ce{TiO2}(110) surface
on top of the Ti(5) atom with the binding energy of 0.14 eV.
The \ce{H-H} bond length of 0.78 \AA\ is slightly increased in comparison with
the free \ce{H2}.  The dissociative state of \ce{H2} on \ce{TiO2}(110) corresponds
to the situation when both H atoms form the OH group with the low coordinated O(2)
bridge atoms on the rutile surface. The binding energy of the dissociated configuration of
\ce{H2} is 1.50 eV; however, the distance between two rows of O(2) atoms
on \ce{TiO2}(110) is 6.50 \AA, which is too large to promote dissociation
of the \ce{H2} adsorbed on top of Ti(5) atom. In this case \ce{H2} would dissociate
in the vicinity of the adsorption center as a first step, followed by
adsorption of H atoms on O(2).
This process would require to overcome a dissociation barrier much higher
than the energy of the molecular adsorption.
In this situation \ce{H2} would likely desorb from the surface and fly
away rather than dissociate.
Therefore, to promote \ce{H2} dissociation on \ce{TiO2}(110), it is necessary
to have adsorption centers on the rutile surface in the vicinity of
low coordinated O(2) atoms. Supported gold clusters can serve as a
source of such centers.

\begin{figure}[htbp]
\begin{center}
\includegraphics[scale=0.6,clip,bb=60 500 550 670]{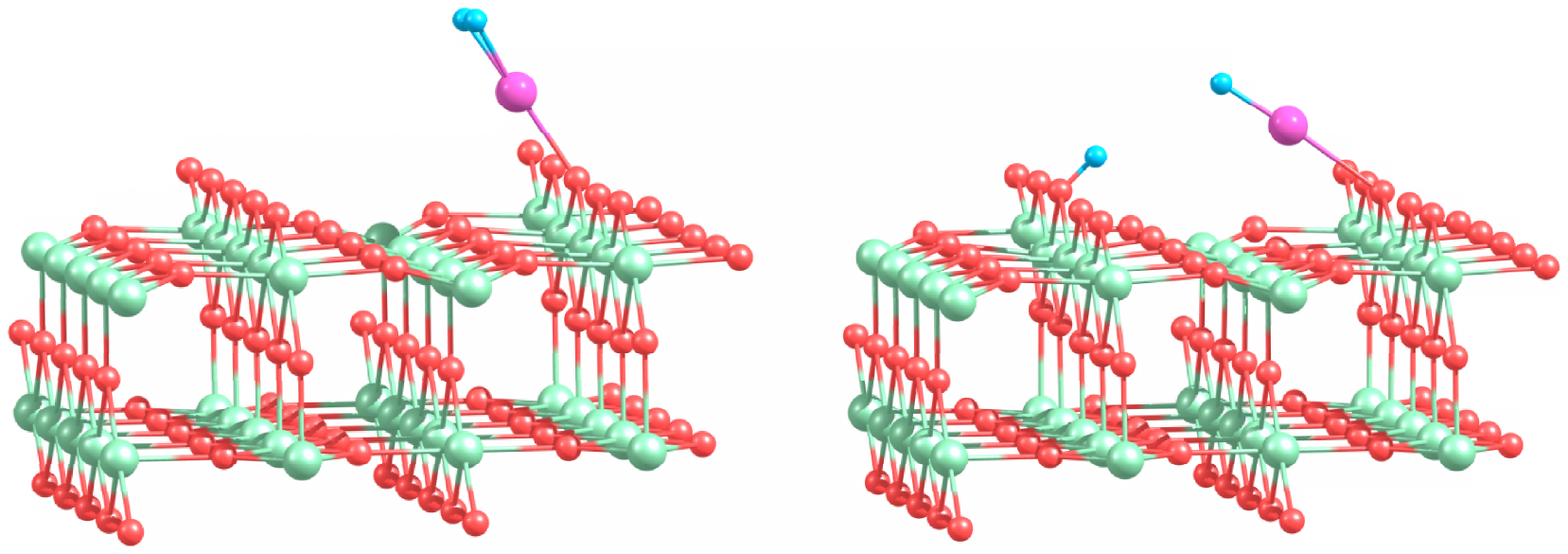}
\end{center}
\caption{Optimized geometries for molecular (left) and dissociative (right)
adsorption of \ce{H2} on Au/\ce{TiO2}.
}
\label{fig:H2_Au1}
\end{figure}

\begin{table}[h]
\small
\centering
  \caption{~Binding energy calculated for molecular,
$E_{b}^{mol}$, and dissociative, $E_{b}^{dis}$,
adsorption of \ce{H2} on \ce{Au_{\it n}}/\ce{TiO2},
and the \ce{H-H} bond length, $r_{\rm H-H}^{mol}$,
in the case of molecular adsorption of \ce{H2}.}
  \label{tbl:binding_supported}
  \begin{tabular}{l c c c}
    \hline
    Cluster           & $r_{\rm H-H}^{mol}$ (\AA) & $E_{b}^{mol}$ (eV) & $E_{b}^{dis}$ (eV)\\
    \hline
    \ce{Au}           &  0.906    & 1.15  & 2.06               \\
    \ce{Au2}          &  0.821    & 0.37  & 0.73$^{(b)}$ 0.90$^{(c)}$ 1.55$^{(d)}$       \\
    \ce{Au8} (2D)     &  0.792    & 0.24  & 0.75$^{(b)}$  1.55$^{(c)}$  \\
    \ce{Au8} (3D)     &  0.753    & 0.09  & 0.55$^{(e)}$  0.95$^{(f)}$ \\
    \ce{Au_{20}} (3D) &  0.795$^{(a)}$ 0.806$^{(a')}$    & 0.17$^{(a)}$  0.13$^{(a')}$  & 0.20$^{(b)}$ 0.83$^{(c)}$ \\
    \ce{Au_{20}} (2D) &  0.798$^{(d)}$ 0.805$^{(d')}$    & 0.11$^{(d)}$  0.11$^{(d')}$  & 0.55$^{(e)}$ 1.42$^{(f)}$\\
    \hline
  \end{tabular}
\end{table}

Fig. \ref{fig:H2_Au1} presents optimized geometries calculated for molecular and dissociative
adsorption of \ce{H2} on Au/\ce{TiO2}.  It is seen from Table \ref{tbl:binding_supported}
that \ce{H2} adsorbs molecularly on Au/\ce{TiO2}
with a binding energy of 1.15 eV, which is considerably larger than the corresponding energy
calculated for \ce{H2} adsorbed on free Au atom.  The adsorbed \ce{H2} is
highly activated, with the \ce{H-H} bond length $r_{\rm H-H}^{mol}$=0.906 \AA.
As a result of \ce{H2} adsorption, the supported Au atom shifts slightly towards the row of
the low coordinated O(2) bridge atoms.  In the previous section we saw that \ce{H2} dissociation
is not favorable on the free Au atom. However, in the case of the supported Au atom,
dissociation of \ce{H2} can occur with formation
of the OH group with the O(2) atom located either in the nearest to Au row of O(2) or
in the next row of O(2), as is shown in Fig. \ref{fig:H2_Au1}. In the latter case
the calculated binding energy $E_{b}^{dis}$=2.06 eV is higher, not only if compared with
the binding energy for molecular adsorption of \ce{H2} on Au/\ce{TiO2}(110), but also
 if compared with the binding energy of the dissociated state of
\ce{H2} on the pure \ce{TiO2}(110) surface.
Thus, we can conclude, from the energetic point of view, that
the \ce{TiO2}(110) support considerably promotes \ce{H2} dissociation on  Au atom.

\begin{figure}[htbp]
\begin{center}
\includegraphics[scale=0.6,clip,bb=60 380 530 710]{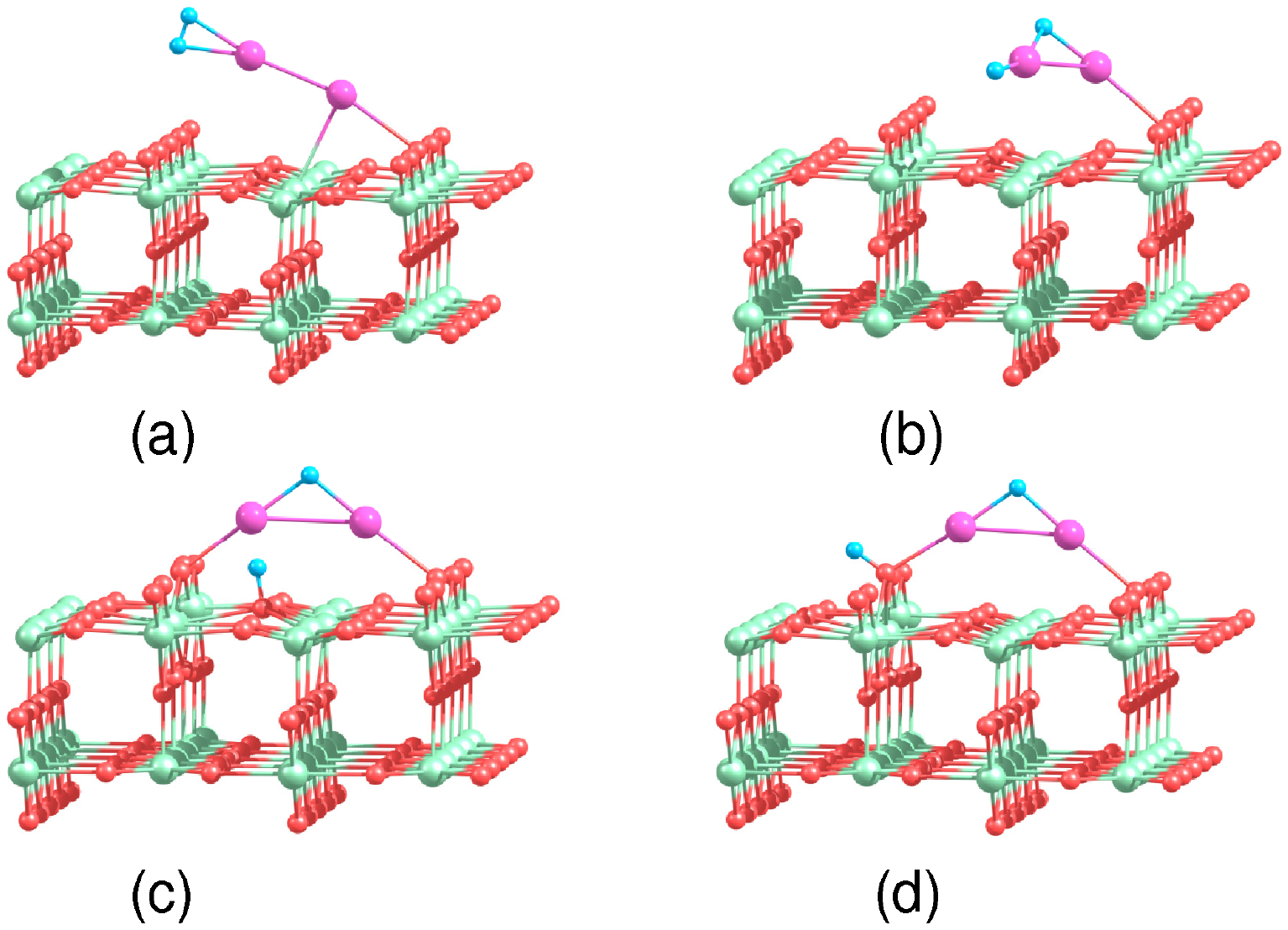}
\end{center}
\caption{(a) Optimized geometry in the case of molecular adsorption of \ce{H2} on \ce{Au2}/\ce{TiO2}.
Optimized geometries in the case of dissociative adsorption of \ce{H2} on \ce{Au2}/\ce{TiO2}:
(b) dissociation of \ce{H2} on supported \ce{Au2};
(c) dissociation of \ce{H2} with formation of the \ce{OH} group on O(3) surface atom;
(d) dissociation of \ce{H2} with formation of the \ce{OH} group on O(2) bridge atom.}
\label{fig:H2_Au2}
\end{figure}

The optimized geometries calculated for molecular and dissociative
adsorption of \ce{H2} on \ce{Au2}/\ce{TiO2} are shown in Fig. \ref{fig:H2_Au2}.
Figs. \ref{fig:H2_Au2}(a) and \ref{fig:H2_Au2}(b) demonstrate that
\ce{H2} binds to the supported \ce{Au2} in a similar way as for the free \ce{Au2}.
However the binding energy calculated for the molecular adsorption
of \ce{H2} on \ce{Au2}/\ce{TiO2} is lower than the corresponding energy obtained for free \ce{Au2},
while the dissociated configuration of \ce{H2} on  \ce{Au2}/\ce{TiO2} is more stable if compared
with dissociative adsorption of \ce{H2} on the free \ce{Au2}. 
Therefore, the interaction of
\ce{Au2} with the support results in the energetic promotion of \ce{H2} dissociation on the
supported \ce{Au2}. We found, however, that the geometry configuration shown
in Fig. \ref{fig:H2_Au2}(b) is not the most stable one for \ce{H2} dissociation.
We found that the hydrogen atom can migrate to the row of O(3) atoms, or to the row of O(2)
bridge atoms on the rutile surface, forming \ce{OH} bonds, as it is shown in
Figs. \ref{fig:H2_Au2}(c) and \ref{fig:H2_Au2}(d), respectively. Hydrogen atom always
binds more strongly with the low coordinated O(2) atoms, as can be seen from
Table \ref{tbl:binding_supported}.

\begin{figure}[htbp]
\begin{center}
\includegraphics[scale=0.6,clip,bb=60 185 530 700]{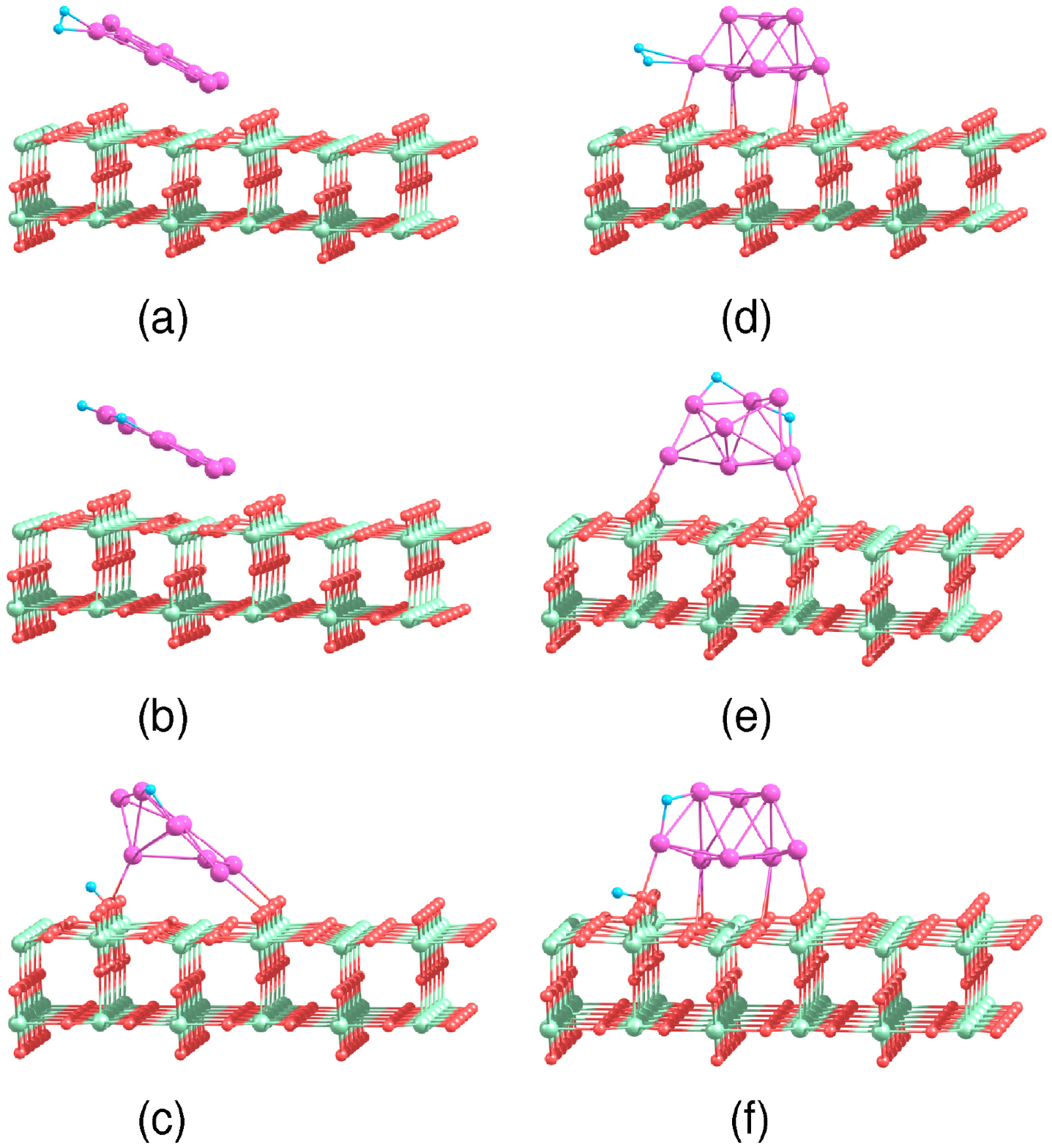}
\end{center}
\caption{Optimized geometries in the case of (a) molecular
adsorption of \ce{H2} on \ce{Au8}(2D)/\ce{TiO2};
dissociative adsorption of \ce{H2} on \ce{Au2}(2D)/\ce{TiO2}:
(b) dissociation of \ce{H2} on supported \ce{Au2}(2D);
(c) dissociation of \ce{H2} with formation of the \ce{OH} group on O(2) bridge atom.
Optimized geometries in the case of (d) molecular adsorption of \ce{H2} on \ce{Au8}(3D)/\ce{TiO2};
dissociative adsorption of \ce{H2} on \ce{Au2}(3D)/\ce{TiO2}:
(e) dissociation of \ce{H2} on supported \ce{Au2}(3D);
(f) dissociation of \ce{H2} with formation of the \ce{OH} group on O(2) bridge atom.}
\label{fig:H2_Au8}
\end{figure}

Figs. \ref{fig:H2_Au8}(a), \ref{fig:H2_Au8}(b), \ref{fig:H2_Au8}(d) and \ref{fig:H2_Au8}(e)
demonstrate that optimized geometries of \ce{H2} in the case of its molecular and dissociative
adsorption on the supported \ce{Au8} (2D) and \ce{Au8} (3D) clusters 
similar to 
those obtained for the corresponding free \ce{Au8} clusters. Note that
the dissociative adsorption of \ce{H2} on \ce{Au8} (3D) results in the considerable
rearrangement of the gold structure as shown in Fig. \ref{fig:H2_Au8}(e).
Tables  \ref{tbl:binding_free} and \ref{tbl:binding_supported}  demonstrate that
the binding energies calculated for molecular adsorption of \ce{H2}
on free and supported \ce{Au8} (2D) and \ce{Au8} (3D)  are quite similar.
However, the dissociative configuration of \ce{H2} on the supported
\ce{Au8} (2D) and \ce{Au8} (3D) clusters is slightly promoted energetically.
Migration of \ce{H} on the low coordinated O(2) atom as it is shown in
Figs. \ref{fig:H2_Au8}(c) and \ref{fig:H2_Au8}(f),
 results in formation of the
\ce{OH} bond and considerable increase in binding energies calculated for \ce{H2} adsorbed
dissociatively. It is seen from Fig. \ref{fig:H2_Au8}(c) that the
planar structure of \ce{Au8} undergoes drastic rearrangements upon \ce{H2} dissociation.
The dissociative adsorption of \ce{H2} on \ce{Au8} is energetically
favorable if compared with the adsorption on the 3D isomer of \ce{Au8}.

\begin{figure}[hptb]
\begin{center}
\includegraphics[scale=0.6,clip,bb=65 90 550 760]{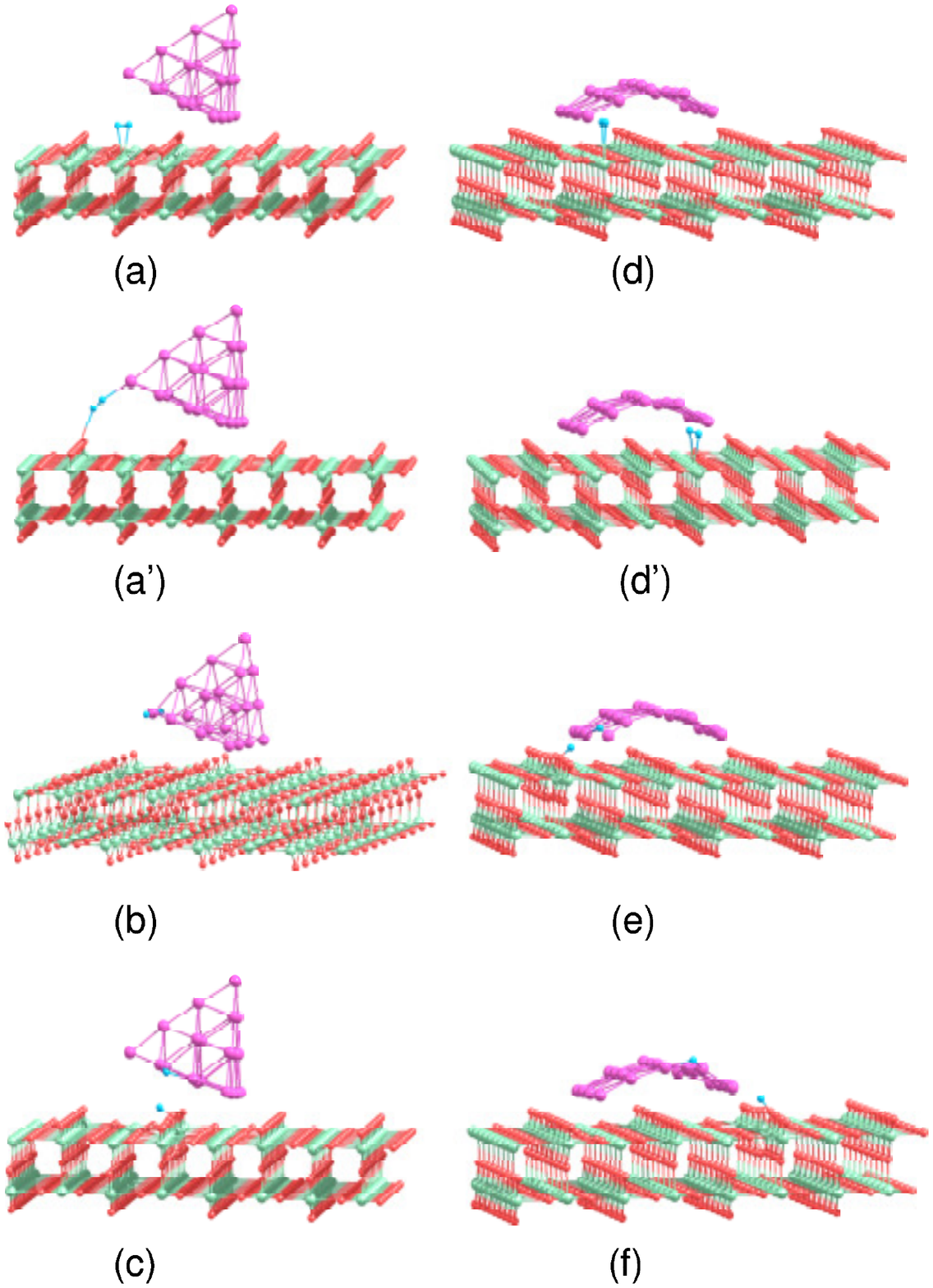}
\end{center}
\caption{Optimized geometries for \ce{H2} adsorbed on \ce{Au_{20}}(3D)/\ce{TiO2} (left) and
\ce{Au_{20}}(2D)/\ce{TiO2} (right).
(a) and (a') Molecular adsorption of \ce{H2} in the vicinity  of \ce{Au_{20}}(3D)/\ce{TiO2};
(b) dissociative adsorption of \ce{H2} at the vertex of \ce{Au_{20}}(3D) with
formation of the \ce{OH} group;
(c) dissociative adsorption of \ce{H2} at edge of \ce{Au_{20}}(3D) with
formation of the \ce{OH} group;
(d) and (d') molecular adsorption of \ce{H2} in the vicinity  of \ce{Au_{20}}(2D)/\ce{TiO2};
(e) dissociative adsorption of \ce{H2} at the corner of \ce{Au_{20}}(2D) with
formation of the \ce{OH} group;
(f) dissociative adsorption of \ce{H2} at edge of \ce{Au_{20}}(2D) with
formation of the \ce{OH} group.}
\label{fig:H2_Au20}
\end{figure}

The \ce{Au_{20}} (3D) and \ce{Au_{20}} (2D)  clusters supported on \ce{TiO2}(110)
are the largest systems studied in the present work. In the previous section it was shown that the
hydrogen molecule binds weakly to the free \ce{Au_{20}} (3D) and \ce{Au_{20}} (2D)
clusters. We have found that, in the case of the supported \ce{Au_{20}}, the
hydrogen molecule adsorbs on top of the Ti(5) atoms on the rutile surface
in the vicinity of \ce{Au_{20}}, rather than directly on the \ce{Au_{20}} clusters,
as it is shown in Figs. \ref{fig:H2_Au20}(a), \ref{fig:H2_Au20}(d) and \ref{fig:H2_Au20}(d').
{The geometry structure when \ce{H2} adsorbs at the vertex of \ce{Au_{20}}(3D), 
bridging Au vertex atom in the cluster and the low coordinated O(2) atom on the rutile
surface is also stable (Fig. \ref{fig:H2_Au20}(a')). } Further \ce{H2} dissociation can
occur either at the vertex  of \ce{Au_{20}} (3D) or at the acute-angled
corner of \ce{Au_{20}} (2D), as is shown in
Fig. \ref{fig:H2_Au20}(b) and Fig. \ref{fig:H2_Au20}(e), respectively.
\ce{H2} dissociation at the acute-angled corner of \ce{Au_{20}} (2D)
is accompanied by severe local structural rearrangement of Au atoms at the
corner of \ce{Au_{20}} (2D) and migration of one H atom to the O(2) atom on the surface.
Note that the dissociation of \ce{H2} at the obtuse-angled corner of the supported
\ce{Au_{20}} (2D) cluster is not energetically favorable,
in comparison to the case of free \ce{Au_{20}} (2D).
The most stable configurations of the dissociated \ce{H2} are those
where one of the \ce{H} atoms binds to the edges
of \ce{Au_{20}} (3D) or \ce{Au_{20}} (2D) that are oriented perpendicularly 
to the rows of O(2) surface atoms,
while another \ce{H} forms the \ce{OH} bond with the O(2) atom on the rutile surface,
as is shown in Figs. \ref{fig:H2_Au20}(c) and \ref{fig:H2_Au20}(f).
We found several geometrical configurations of this type with binding energies of 1.19 - 1.42 eV.
Dissociation at the edge of the supported 2D isomer of \ce{Au_{20}} is always energetically
favorable.

Thus, combination (interplay) of several factors such as geometry structure,
cluster dimensionality, presence of the low coordinated oxygen atoms
in the vicinity of the cluster-surface interface, etc.
are important for \ce{H2} dissociation.

It is well known that, in the case of catalytic oxidation reactions by
molecular oxygen on free and supported gold nanoparticles, the charge transfer
from the gold to the antibonding orbital of \ce{O2} is responsible for the
catalytic activation and dissociation of \ce{O2}. However, in the case of
\ce{H2} dissociation the  analysis of the Bader charges\cite{Bader90,Henkelman06}
demonstrate that there is no considerable charge transfer
between the adsorbed hydrogen and the gold clusters (free or supported).
We found the largest charge transfer occurs for \ce{H2} adsorbed molecularly on Au/\ce{TiO2}
(Fig.  \ref{fig:H2_Au1}). In this case the calculated Bader charge localized
on \ce{H2} is +0.11{e}, where {e} is the elementary charge.
Although such a charge transfer is relatively small, it might be responsible for the strong
enlargement of the \ce{H-H} bond length in \ce{H2-Au}/\ce{TiO2} up to 0.906 \AA.
Nevertheless, we have not found the direct correlation 
of the hydrogen dissociation with the charge transfer between hydrogen and gold atoms.
On the other hand, formation of the \ce{OH}
group with the bridge O(2) atom on the rutile surface is accompanied by the large
charge transfer from H atom to O(2), resulting in a Bader net charge of H in \ce{OH} equal to
+0.75{e}.
Thus, we can conclude that the catalytic activity of gold nanoparticles for \ce{O2} dissociation
would depend on the electronic structure and the size
of the nanoparticles; however in the case of \ce{H2} dissociation
it will be proportional to the number of gold atoms located in the vicinity
of the low coordinated O(2) atoms at the nanoparticle -- surface interface.

\section{Conclusions}

The present theoretical study demonstrates
that adsorption of \ce{H2} on \ce{Au_{\it n}} (n= 1, 2, 8, 20)
strongly depends on cluster size, geometry structure, flexibility and
interaction with the support material.
Strong interaction with the support can stabilize 2D structures of gold
clusters on the surface.
This is important finding, because
the energy release due to
\ce{H2} dissociation is the largest for the 2D isomers of gold clusters.
Rutile \ce{TiO2}(110) support energetically promotes \ce{H2}
dissociation on gold clusters.
The formation of the \ce{OH} group near the supported gold cluster
is an important condition for \ce{H2} dissociation.
We have shown that the active sites towards \ce{H2} dissociation on the supported
\ce{Au_{\it n}} are located at corners and edges of the gold cluster in the
vicinity of the low coordinated oxygen atoms on \ce{TiO2}(110).
Therefore catalytic activity of a gold nanoparticle supported on the rutile \ce{TiO2}(110)
surface is proportional to the length of the perimeter interface between the nanoparticle
and the support, in accordance with the recent experimental findings.\cite{Fujitani09}

In the present work we have systematically studied adsorption and dissociation of \ce{H2} based
on the assumption of the total energy criterion.
Thus, we have identified a number of the
most stable configurations for molecular and dissociative states of \ce{H2}
adsorbed on free and supported gold nanoparticles.
Conditions leading to the stabilization of the
dissociated state of \ce{H2} have been disclosed.

However, many interesting and important questions regarding the mechanism
of \ce{H2} dissociation on \ce{TiO2}(110) supported \ce{Au_{\it n}} clusters 
remain unanswered. Thus, it is necessary to elucidate the full reaction pathways
and to estimate the reaction barriers for \ce{H2} dissociation. This will allow favorable
reaction channels and reaction dynamics to be identified.
As a next step one should consider simple hydrogenation reactions, such as, 
for example, hydrogen peroxide formation from \ce{H2} and \ce{O2} 
catalyzed by the supported gold clusters.  
We can suggest that structures active for \ce{H2} dissociation, 
but not active enough for \ce{O2} dissociation could work as catalysts 
for \ce{H2O2} synthesis.

In many cases we have found an ensemble of initial and final
configurations of the adsorbed hydrogen with small differences in binding energies,
but possible difference in the reaction barriers. 
Accounting for the geometry of the initial and final states
can lead to further non-equivalence of different reaction pathways
and may have an important impact on the change of system entropy.
It has been argued\cite{Brechignac_PRL96,Brechignac_PRL98,Obolensky05} that
accounting for an entropy change contribution to the free energy
of the system is necessary for correct
description of the temperature dependencies
of the branching ratios between different reaction channels,
while purely energetic considerations based on reaction barrier
heights can fail.

Finally, we have considered the pure, defect free rutile \ce{TiO2}(110) surface.
It is well known that the surface defects, especially O vacancies
in metal-oxide supports, or extra O atoms adsorbed on the surface
can change drastically the catalytic activity of the supported
gold clusters in oxidation reactions by molecular oxygen.\cite{Sanchez99,Yoon05}
{Interaction of small gold clusters with a partially reduced rutile 
\ce{TiO2}(110) support can be rather complicated; supported gold clusters 
can either receive or donate electrons to the substrate 
depending on the cluster size.\cite{Chretien07}}
The role of the surface
defects in the process of \ce{H2} dissociation is not so obvious and requires
additional systematic investigation.
Such defects, in particular additional O atoms adsorbed on the surface
in the vicinity of the cluster -- surface interface can
probably play a role in \ce{H2} dissociation and hence can modify the cluster reactivity.
Understanding how to control such chemical reactions on
a cluster surface is a vital task for nanocatalysis.

We hope that the present discussion will stimulate further experimental and theoretical
investigations of the processes considered.

%
%
%

\section*{Acknowledgements}
This work was supported by the Global COE Program
(Project No. B01:  Catalysis as the Basis for Innovation
in Materials Science) from the Ministry of Education,
Culture, Sports, Science and Technology, Japan;
the Grant-in-Aid for the Project on Strategic
Utilization of Elements
and the JSPS Grant-in-Aid for Scientific Research C.
The computations were performed using the Research Center for
Computational Science, Okazaki, Japan.




\footnotesize{
\bibliography{Au_catalysis} 

\providecommand*{\mcitethebibliography}{\thebibliography}
\csname @ifundefined\endcsname{endmcitethebibliography}
{\let\endmcitethebibliography\endthebibliography}{}
\begin{mcitethebibliography}{97}
\providecommand*{\natexlab}[1]{#1}
\providecommand*{\mciteSetBstSublistMode}[1]{}
\providecommand*{\mciteSetBstMaxWidthForm}[2]{}
\providecommand*{\mciteBstWouldAddEndPuncttrue}
  {\def\EndOfBibitem{\unskip.}}
\providecommand*{\mciteBstWouldAddEndPunctfalse}
  {\let\EndOfBibitem\relax}
\providecommand*{\mciteSetBstMidEndSepPunct}[3]{}
\providecommand*{\mciteSetBstSublistLabelBeginEnd}[3]{}
\providecommand*{\EndOfBibitem}{}
\mciteSetBstSublistMode{f}
\mciteSetBstMaxWidthForm{subitem}
{(\emph{\alph{mcitesubitemcount}})}
\mciteSetBstSublistLabelBeginEnd{\mcitemaxwidthsubitemform\space}
{\relax}{\relax}

\bibitem[Haruta \emph{et~al.}(1987)Haruta, Kobayashi, Sano, and
  Yamada]{Haruta87}
M.~Haruta, T.~Kobayashi, H.~Sano and N.~Yamada, \emph{Chem. Lett.}, 1987,
  \textbf{16}, 405--408\relax
\mciteBstWouldAddEndPuncttrue
\mciteSetBstMidEndSepPunct{\mcitedefaultmidpunct}
{\mcitedefaultendpunct}{\mcitedefaultseppunct}\relax
\EndOfBibitem
\bibitem[Haruta(2003)]{Haruta03}
M.~Haruta, \emph{The Chemical Record}, 2003, \textbf{3}, 75--87\relax
\mciteBstWouldAddEndPuncttrue
\mciteSetBstMidEndSepPunct{\mcitedefaultmidpunct}
{\mcitedefaultendpunct}{\mcitedefaultseppunct}\relax
\EndOfBibitem
\bibitem[Thompson(2007)]{Thompson07}
D.~T. Thompson, \emph{Nanotaday}, 2007, \textbf{2}, 40--43\relax
\mciteBstWouldAddEndPuncttrue
\mciteSetBstMidEndSepPunct{\mcitedefaultmidpunct}
{\mcitedefaultendpunct}{\mcitedefaultseppunct}\relax
\EndOfBibitem
\bibitem[Haruta(1997)]{Haruta97}
M.~Haruta, \emph{Catalysis Today}, 1997, \textbf{36}, 153--166\relax
\mciteBstWouldAddEndPuncttrue
\mciteSetBstMidEndSepPunct{\mcitedefaultmidpunct}
{\mcitedefaultendpunct}{\mcitedefaultseppunct}\relax
\EndOfBibitem
\bibitem[Turner \emph{et~al.}(2008)Turner, Golovko, Vaughan, Abdulkin,
  Berenguer-Murcia, Tikhov, Johnson, and Lambert]{Turner08}
M.~Turner, V.~B. Golovko, O.~P.~H. Vaughan, P.~Abdulkin, A.~Berenguer-Murcia,
  M.~S. Tikhov, B.~F.~G. Johnson and R.~M. Lambert, \emph{Nature}, 2008,
  \textbf{454}, 981--984\relax
\mciteBstWouldAddEndPuncttrue
\mciteSetBstMidEndSepPunct{\mcitedefaultmidpunct}
{\mcitedefaultendpunct}{\mcitedefaultseppunct}\relax
\EndOfBibitem
\bibitem[Hughes \emph{et~al.}(2005)Hughes, Xu, Jenkins, McMorn, Landon, Enache,
  Carley, Attard, Hutchings, King, Stitt, Johnston, Griffin, and
  Kiely]{Hughes05}
M.~D. Hughes, Y.-J. Xu, P.~Jenkins, P.~McMorn, P.~Landon, D.~I. Enache, A.~F.
  Carley, G.~A. Attard, G.~J. Hutchings, F.~King, E.~H. Stitt, P.~Johnston,
  K.~Griffin and C.~J. Kiely, \emph{Nature}, 2005, \textbf{437},
  1132--1135\relax
\mciteBstWouldAddEndPuncttrue
\mciteSetBstMidEndSepPunct{\mcitedefaultmidpunct}
{\mcitedefaultendpunct}{\mcitedefaultseppunct}\relax
\EndOfBibitem
\bibitem[Tsunoyama \emph{et~al.}(2005)Tsunoyama, Sakurai, Negishi, and
  Tsukuda]{Tsunoyama05}
H.~Tsunoyama, H.~Sakurai, Y.~Negishi and T.~Tsukuda, \emph{J. Am. Chem. Soc.},
  2005, \textbf{127}, 9374--9375\relax
\mciteBstWouldAddEndPuncttrue
\mciteSetBstMidEndSepPunct{\mcitedefaultmidpunct}
{\mcitedefaultendpunct}{\mcitedefaultseppunct}\relax
\EndOfBibitem
\bibitem[Landon \emph{et~al.}(2002)Landon, Collier, Papworth, Kiely, and
  Hutchings]{Landon02}
P.~Landon, P.~J. Collier, A.~J. Papworth, C.~J. Kiely and G.~Hutchings,
  \emph{Chem. Commun.}, 2002,  2058--2059\relax
\mciteBstWouldAddEndPuncttrue
\mciteSetBstMidEndSepPunct{\mcitedefaultmidpunct}
{\mcitedefaultendpunct}{\mcitedefaultseppunct}\relax
\EndOfBibitem
\bibitem[Lyalin and Taketsugu(2009)]{Lyalin09}
A.~Lyalin and T.~Taketsugu, \emph{J. Phys. Chem. C}, 2009, \textbf{113},
  12930--12934\relax
\mciteBstWouldAddEndPuncttrue
\mciteSetBstMidEndSepPunct{\mcitedefaultmidpunct}
{\mcitedefaultendpunct}{\mcitedefaultseppunct}\relax
\EndOfBibitem
\bibitem[Lyalin and Taketsugu(2010)]{Lyalin10a}
A.~Lyalin and T.~Taketsugu, \emph{J. Phys. Chem. Lett.}, 2010, \textbf{1},
  1752--1757\relax
\mciteBstWouldAddEndPuncttrue
\mciteSetBstMidEndSepPunct{\mcitedefaultmidpunct}
{\mcitedefaultendpunct}{\mcitedefaultseppunct}\relax
\EndOfBibitem
\bibitem[Herzing \emph{et~al.}(2008)Herzing, Kiely, Carley, Landon, and
  Hutchings]{Herzing08}
A.~A. Herzing, C.~J. Kiely, A.~F. Carley, P.~Landon and G.~J. Hutchings,
  \emph{Science}, 2008, \textbf{321}, 1331--1335\relax
\mciteBstWouldAddEndPuncttrue
\mciteSetBstMidEndSepPunct{\mcitedefaultmidpunct}
{\mcitedefaultendpunct}{\mcitedefaultseppunct}\relax
\EndOfBibitem
\bibitem[Rodr\'{i}guez-V\'{a}zquez
  \emph{et~al.}(2008)Rodr\'{i}guez-V\'{a}zquez, Blanco, Lourido,
  V\'{a}zquez-V\'{a}zquez, Pastor, Planes, Rivas, and
  L\'{o}pez-Quintela]{Maria08}
M.~J. Rodr\'{i}guez-V\'{a}zquez, M.~C. Blanco, R.~Lourido,
  C.~V\'{a}zquez-V\'{a}zquez, E.~Pastor, G.~A. Planes, J.~Rivas and M.~A.
  L\'{o}pez-Quintela, \emph{Langmuir}, 2008, \textbf{24}, 12690--12694\relax
\mciteBstWouldAddEndPuncttrue
\mciteSetBstMidEndSepPunct{\mcitedefaultmidpunct}
{\mcitedefaultendpunct}{\mcitedefaultseppunct}\relax
\EndOfBibitem
\bibitem[Heiz and Landman(2007)]{Nanocat07}
\emph{Nanocatalysis}, ed. U.~Heiz and U.~Landman, Springer, Berlin, Heidelberg,
  New York, 2007\relax
\mciteBstWouldAddEndPuncttrue
\mciteSetBstMidEndSepPunct{\mcitedefaultmidpunct}
{\mcitedefaultendpunct}{\mcitedefaultseppunct}\relax
\EndOfBibitem
\bibitem[Hvolb{\ae}k \emph{et~al.}(2007)Hvolb{\ae}k, Janssens, Clausen, Falsig,
  Christensen, and N{\o}rskov]{Hvolbaek07}
B.~Hvolb{\ae}k, T.~V.~W. Janssens, B.~S. Clausen, H.~Falsig, C.~H. Christensen
  and J.~K. N{\o}rskov, \emph{Nanotoday}, 2007, \textbf{2}, 14--18\relax
\mciteBstWouldAddEndPuncttrue
\mciteSetBstMidEndSepPunct{\mcitedefaultmidpunct}
{\mcitedefaultendpunct}{\mcitedefaultseppunct}\relax
\EndOfBibitem
\bibitem[Lang \emph{et~al.}(2009)Lang, Bernhardt, Barnett, Yoon, and
  Landman]{Lang09}
S.~M. Lang, T.~M. Bernhardt, R.~N. Barnett, B.~Yoon and U.~Landman, \emph{J.
  Am. Chem. Soc.}, 2009, \textbf{131}, 8939--8951\relax
\mciteBstWouldAddEndPuncttrue
\mciteSetBstMidEndSepPunct{\mcitedefaultmidpunct}
{\mcitedefaultendpunct}{\mcitedefaultseppunct}\relax
\EndOfBibitem
\bibitem[Landman \emph{et~al.}(2007)Landman, Yoon, Zhang, Heiz, and
  Arenz]{Landman07}
U.~Landman, B.~Yoon, C.~Zhang, U.~Heiz and M.~Arenz, \emph{Topics in
  Catalisis}, 2007, \textbf{44}, 145--158\relax
\mciteBstWouldAddEndPuncttrue
\mciteSetBstMidEndSepPunct{\mcitedefaultmidpunct}
{\mcitedefaultendpunct}{\mcitedefaultseppunct}\relax
\EndOfBibitem
\bibitem[Harding \emph{et~al.}(2009)Harding, Habibpour, Kunz, Farnbacher, Heiz,
  Yoon, and Landman]{Harding09}
C.~Harding, V.~Habibpour, S.~Kunz, A.~N.-S. Farnbacher, U.~Heiz, B.~Yoon and
  U.~Landman, \emph{J. Am. Chem. Soc.}, 2009, \textbf{131}, 538--548\relax
\mciteBstWouldAddEndPuncttrue
\mciteSetBstMidEndSepPunct{\mcitedefaultmidpunct}
{\mcitedefaultendpunct}{\mcitedefaultseppunct}\relax
\EndOfBibitem
\bibitem[Rodr\'{i}guez \emph{et~al.}(2010)Rodr\'{i}guez, Feria, Jirsak,
  Takahashi, Nakamura, and Illas]{Rodriguez10}
J.~A. Rodr\'{i}guez, L.~Feria, T.~Jirsak, Y.~Takahashi, K.~Nakamura and
  F.~Illas, \emph{J. Am. Chem. Soc.}, 2010, \textbf{132}, 3177--3186\relax
\mciteBstWouldAddEndPuncttrue
\mciteSetBstMidEndSepPunct{\mcitedefaultmidpunct}
{\mcitedefaultendpunct}{\mcitedefaultseppunct}\relax
\EndOfBibitem
\bibitem[Coquet \emph{et~al.}(2008)Coquet, Howard, and Willock]{Coquet08}
R.~Coquet, K.~L. Howard and D.~J. Willock, \emph{Chem. Soc. Rev.}, 2008,
  \textbf{37}, 2046--2076\relax
\mciteBstWouldAddEndPuncttrue
\mciteSetBstMidEndSepPunct{\mcitedefaultmidpunct}
{\mcitedefaultendpunct}{\mcitedefaultseppunct}\relax
\EndOfBibitem
\bibitem[Sanchez \emph{et~al.}(1999)Sanchez, Abbet, Heiz, Schneider, Hakkinen,
  Barnett, and Landman]{Sanchez99}
A.~Sanchez, S.~Abbet, U.~Heiz, W.-D. Schneider, H.~Hakkinen, R.~N. Barnett and
  U.~Landman, \emph{J. Phys. Chem. A}, 1999, \textbf{103}, 9573--9578\relax
\mciteBstWouldAddEndPuncttrue
\mciteSetBstMidEndSepPunct{\mcitedefaultmidpunct}
{\mcitedefaultendpunct}{\mcitedefaultseppunct}\relax
\EndOfBibitem
\bibitem[Dat\'{e} \emph{et~al.}(2004)Dat\'{e}, Okumura, Tsubota, and
  Haruta]{Date04}
M.~Dat\'{e}, M.~Okumura, S.~Tsubota and M.~Haruta, \emph{Angew. Chem. Int.
  Ed.}, 2004, \textbf{43}, 2129--2132\relax
\mciteBstWouldAddEndPuncttrue
\mciteSetBstMidEndSepPunct{\mcitedefaultmidpunct}
{\mcitedefaultendpunct}{\mcitedefaultseppunct}\relax
\EndOfBibitem
\bibitem[Matthey \emph{et~al.}(2007)Matthey, Wang, Wendt, Matthiesen, Schaub,
  L{\ae}gsgaard, Hammer, and Besenbacher]{Matthey07}
D.~Matthey, J.~G. Wang, S.~Wendt, J.~Matthiesen, R.~Schaub, E.~L{\ae}gsgaard,
  B.~Hammer and F.~Besenbacher, \emph{Science}, 2007, \textbf{315},
  1692--1696\relax
\mciteBstWouldAddEndPuncttrue
\mciteSetBstMidEndSepPunct{\mcitedefaultmidpunct}
{\mcitedefaultendpunct}{\mcitedefaultseppunct}\relax
\EndOfBibitem
\bibitem[Joshi \emph{et~al.}(2006)Joshi, Delgass, and Thomson]{Joshi06}
A.~M. Joshi, W.~N. Delgass and K.~T. Thomson, \emph{J. Phys. Chem. B}, 2006,
  \textbf{110}, 23373--23387\relax
\mciteBstWouldAddEndPuncttrue
\mciteSetBstMidEndSepPunct{\mcitedefaultmidpunct}
{\mcitedefaultendpunct}{\mcitedefaultseppunct}\relax
\EndOfBibitem
\bibitem[Lyalin and Taketsugu(2010)]{Lyalin10}
A.~Lyalin and T.~Taketsugu, \emph{J. Phys. Chem. C}, 2010, \textbf{114},
  2484--2493\relax
\mciteBstWouldAddEndPuncttrue
\mciteSetBstMidEndSepPunct{\mcitedefaultmidpunct}
{\mcitedefaultendpunct}{\mcitedefaultseppunct}\relax
\EndOfBibitem
\bibitem[Jia \emph{et~al.}(2000)Jia, Haraki, Kondo, Domen, and Tamaru]{Jia00}
J.~Jia, K.~Haraki, J.~N. Kondo, K.~Domen and K.~Tamaru, \emph{J. Phys. Chem.
  B}, 2000, \textbf{104}, 11153--11156\relax
\mciteBstWouldAddEndPuncttrue
\mciteSetBstMidEndSepPunct{\mcitedefaultmidpunct}
{\mcitedefaultendpunct}{\mcitedefaultseppunct}\relax
\EndOfBibitem
\bibitem[Choudhary \emph{et~al.}(2003)Choudhary, Sivadinarayana, Datye, Kumar,
  and Goodman]{Choudhary03}
T.~V. Choudhary, C.~Sivadinarayana, A.~K. Datye, D.~Kumar and D.~W. Goodman,
  \emph{Catal. Lett.}, 2003, \textbf{86}, 1\relax
\mciteBstWouldAddEndPuncttrue
\mciteSetBstMidEndSepPunct{\mcitedefaultmidpunct}
{\mcitedefaultendpunct}{\mcitedefaultseppunct}\relax
\EndOfBibitem
\bibitem[Bailie and Hutchings(1999)]{Bailie99}
J.~E. Bailie and G.~J. Hutchings, \emph{Chem. Commun.}, 1999,  2151--2152\relax
\mciteBstWouldAddEndPuncttrue
\mciteSetBstMidEndSepPunct{\mcitedefaultmidpunct}
{\mcitedefaultendpunct}{\mcitedefaultseppunct}\relax
\EndOfBibitem
\bibitem[Schimpf \emph{et~al.}(2002)Schimpf, Martin~Lucas, Mohra, Rodemerck,
  Br\"{u}ckner, Radnik, Hofmeister, and Claus]{Schimpf02}
S.~Schimpf, M.~Martin~Lucas, C.~Mohra, U.~Rodemerck, A.~Br\"{u}ckner,
  J.~Radnik, H.~Hofmeister and P.~Claus, \emph{Catalysis Today}, 2002,
  \textbf{72}, 63--78\relax
\mciteBstWouldAddEndPuncttrue
\mciteSetBstMidEndSepPunct{\mcitedefaultmidpunct}
{\mcitedefaultendpunct}{\mcitedefaultseppunct}\relax
\EndOfBibitem
\bibitem[Okumura \emph{et~al.}(2002)Okumura, Akita, and Haruta]{Okumura02}
M.~Okumura, T.~Akita and M.~Haruta, \emph{Catalysis Today}, 2002, \textbf{74},
  265--269\relax
\mciteBstWouldAddEndPuncttrue
\mciteSetBstMidEndSepPunct{\mcitedefaultmidpunct}
{\mcitedefaultendpunct}{\mcitedefaultseppunct}\relax
\EndOfBibitem
\bibitem[Mohr \emph{et~al.}(2003)Mohr, Hofmeister, Radnik, and Claus]{Mohr03}
C.~Mohr, H.~Hofmeister, J.~Radnik and P.~Claus, \emph{J. Am. Chem. Soc.}, 2003,
  \textbf{125}, 1905--1911\relax
\mciteBstWouldAddEndPuncttrue
\mciteSetBstMidEndSepPunct{\mcitedefaultmidpunct}
{\mcitedefaultendpunct}{\mcitedefaultseppunct}\relax
\EndOfBibitem
\bibitem[Zanella \emph{et~al.}(2004)Zanella, Louis, Giorgio, and
  Touroude]{Zanella04}
R.~Zanella, C.~Louis, S.~Giorgio and R.~Touroude, \emph{J. Catal.}, 2004,
  \textbf{223}, 328--339\relax
\mciteBstWouldAddEndPuncttrue
\mciteSetBstMidEndSepPunct{\mcitedefaultmidpunct}
{\mcitedefaultendpunct}{\mcitedefaultseppunct}\relax
\EndOfBibitem
\bibitem[Claus(2005)]{Claus05}
P.~Claus, \emph{Appl. Catal. A: Gen.}, 2005, \textbf{291}, 222--229\relax
\mciteBstWouldAddEndPuncttrue
\mciteSetBstMidEndSepPunct{\mcitedefaultmidpunct}
{\mcitedefaultendpunct}{\mcitedefaultseppunct}\relax
\EndOfBibitem
\bibitem[Zhu \emph{et~al.}(2010)Zhu, Qian, Drake, and Jin]{Zhu10}
Y.~Zhu, H.~Qian, B.~A. Drake and R.~Jin, \emph{Angew. Chem. Int. Ed.}, 2010,
  \textbf{49}, 1295--1298\relax
\mciteBstWouldAddEndPuncttrue
\mciteSetBstMidEndSepPunct{\mcitedefaultmidpunct}
{\mcitedefaultendpunct}{\mcitedefaultseppunct}\relax
\EndOfBibitem
\bibitem[Zhu \emph{et~al.}(2010)Zhu, Wu, Gayathri, Qian, Gil, and Jin]{Zhu10a}
Y.~Zhu, Z.~Wu, C.~Gayathri, H.~Qian, R.~R. Gil and R.~Jin, \emph{J. Catal.},
  2010, \textbf{271}, 155--160\relax
\mciteBstWouldAddEndPuncttrue
\mciteSetBstMidEndSepPunct{\mcitedefaultmidpunct}
{\mcitedefaultendpunct}{\mcitedefaultseppunct}\relax
\EndOfBibitem
\bibitem[Bus \emph{et~al.}(2005)Bus, Miller, and van Bokhoven]{Bus05}
E.~Bus, J.~T. Miller and J.~A. van Bokhoven, \emph{J. Phys. Chem. B}, 2005,
  \textbf{109}, 14581--14587\relax
\mciteBstWouldAddEndPuncttrue
\mciteSetBstMidEndSepPunct{\mcitedefaultmidpunct}
{\mcitedefaultendpunct}{\mcitedefaultseppunct}\relax
\EndOfBibitem
\bibitem[Mohr \emph{et~al.}(2003)Mohr, Hofmeister, and Claus]{Mohr03a}
C.~Mohr, H.~Hofmeister and P.~Claus, \emph{J. Catal.}, 2003, \textbf{213},
  86--94\relax
\mciteBstWouldAddEndPuncttrue
\mciteSetBstMidEndSepPunct{\mcitedefaultmidpunct}
{\mcitedefaultendpunct}{\mcitedefaultseppunct}\relax
\EndOfBibitem
\bibitem[Fujitani \emph{et~al.}(2009)Fujitani, Nakamura, Akita, Okumura, and
  Haruta]{Fujitani09}
T.~Fujitani, I.~Nakamura, T.~Akita, M.~Okumura and M.~Haruta, \emph{Angew.
  Chem. Int. Ed.}, 2009, \textbf{48}, 9515--9518\relax
\mciteBstWouldAddEndPuncttrue
\mciteSetBstMidEndSepPunct{\mcitedefaultmidpunct}
{\mcitedefaultendpunct}{\mcitedefaultseppunct}\relax
\EndOfBibitem
\bibitem[Andrews \emph{et~al.}(2004)Andrews, Xuefeng~Wang, Manceron, and
  Balasubramanian]{Andrews04}
L.~Andrews, X.~Xuefeng~Wang, L.~Manceron and K.~Balasubramanian, \emph{J. Phys.
  Chem. A}, 2004, \textbf{108}, 2936--2940\relax
\mciteBstWouldAddEndPuncttrue
\mciteSetBstMidEndSepPunct{\mcitedefaultmidpunct}
{\mcitedefaultendpunct}{\mcitedefaultseppunct}\relax
\EndOfBibitem
\bibitem[Andrews(2004)]{Andrews04a}
L.~Andrews, \emph{Chem. Soc. Rev.}, 2004, \textbf{33}, 123--132\relax
\mciteBstWouldAddEndPuncttrue
\mciteSetBstMidEndSepPunct{\mcitedefaultmidpunct}
{\mcitedefaultendpunct}{\mcitedefaultseppunct}\relax
\EndOfBibitem
\bibitem[Zanchet \emph{et~al.}(2010)Zanchet, Roncero, Omar, Paniagua, and
  Aguado]{Zanchet10}
A.~Zanchet, O.~Roncero, S.~Omar, M.~Paniagua and A.~Aguado, \emph{J. Chem.
  Phys.}, 2010, \textbf{132}, 034301/1--10\relax
\mciteBstWouldAddEndPuncttrue
\mciteSetBstMidEndSepPunct{\mcitedefaultmidpunct}
{\mcitedefaultendpunct}{\mcitedefaultseppunct}\relax
\EndOfBibitem
\bibitem[Varganov \emph{et~al.}(2004)Varganov, Olson, Gordon, Mills, and
  Metiu]{Varganov04}
S.~A. Varganov, R.~M. Olson, M.~S. Gordon, G.~Mills and H.~Metiu, \emph{J.
  Chem. Phys.}, 2004, \textbf{120}, 5169--5175\relax
\mciteBstWouldAddEndPuncttrue
\mciteSetBstMidEndSepPunct{\mcitedefaultmidpunct}
{\mcitedefaultendpunct}{\mcitedefaultseppunct}\relax
\EndOfBibitem
\bibitem[Okumura \emph{et~al.}(2005)Okumura, Kitagawa, Haruta, and
  Yamaguchi]{Okumura05}
M.~Okumura, Y.~Kitagawa, M.~Haruta and K.~Yamaguchi, \emph{Appl. Catal. A:
  Gen.}, 2005, \textbf{291}, 37--44\relax
\mciteBstWouldAddEndPuncttrue
\mciteSetBstMidEndSepPunct{\mcitedefaultmidpunct}
{\mcitedefaultendpunct}{\mcitedefaultseppunct}\relax
\EndOfBibitem
\bibitem[Barrio \emph{et~al.}(2006)Barrio, Liu, Rodr\'{i}guez,
  Campos-Mart\'{i}n, and Fierro]{Barrio06}
L.~Barrio, P.~Liu, J.~A. Rodr\'{i}guez, J.~M. Campos-Mart\'{i}n and J.~L.~G.
  Fierro, \emph{J. Chem. Phys.}, 2006, \textbf{125}, 164715/1--5\relax
\mciteBstWouldAddEndPuncttrue
\mciteSetBstMidEndSepPunct{\mcitedefaultmidpunct}
{\mcitedefaultendpunct}{\mcitedefaultseppunct}\relax
\EndOfBibitem
\bibitem[Ghebriel and Kshirsagar(2007)]{Ghebriel07}
H.~W. Ghebriel and A.~Kshirsagar, \emph{J. Chem. Phys.}, 2007, \textbf{126},
  244705/1--9\relax
\mciteBstWouldAddEndPuncttrue
\mciteSetBstMidEndSepPunct{\mcitedefaultmidpunct}
{\mcitedefaultendpunct}{\mcitedefaultseppunct}\relax
\EndOfBibitem
\bibitem[Joshi \emph{et~al.}(2007)Joshi, Delgass, and Thomson]{Joshi07}
A.~M. Joshi, W.~N. Delgass and K.~T. Thomson, \emph{Topics in Catalisis}, 2007,
  \textbf{44}, 27--39\relax
\mciteBstWouldAddEndPuncttrue
\mciteSetBstMidEndSepPunct{\mcitedefaultmidpunct}
{\mcitedefaultendpunct}{\mcitedefaultseppunct}\relax
\EndOfBibitem
\bibitem[Corma \emph{et~al.}(2007)Corma, Boronat, Gonz\'{a}lez, and
  Illas]{Corma07}
A.~Corma, M.~Boronat, S.~Gonz\'{a}lez and F.~Illas, \emph{Chem. Commun.}, 2007,
   3371--3373\relax
\mciteBstWouldAddEndPuncttrue
\mciteSetBstMidEndSepPunct{\mcitedefaultmidpunct}
{\mcitedefaultendpunct}{\mcitedefaultseppunct}\relax
\EndOfBibitem
\bibitem[Wells \emph{et~al.}(2004)Wells, Delgass, and Thomson]{Wells04}
D.~H. Wells, Jr., W.~N. Delgass and K.~T. Thomson, \emph{J. Catal.}, 2004,
  \textbf{225}, 69--77\relax
\mciteBstWouldAddEndPuncttrue
\mciteSetBstMidEndSepPunct{\mcitedefaultmidpunct}
{\mcitedefaultendpunct}{\mcitedefaultseppunct}\relax
\EndOfBibitem
\bibitem[Kacprzak \emph{et~al.}(2009)Kacprzak, Akola, and
  H\"{a}kkinen]{Kacprzak09}
K.~A. Kacprzak, J.~Akola and H.~H\"{a}kkinen, \emph{Phys. Chem. Chem. Phys.},
  2009, \textbf{11}, 6359--6364\relax
\mciteBstWouldAddEndPuncttrue
\mciteSetBstMidEndSepPunct{\mcitedefaultmidpunct}
{\mcitedefaultendpunct}{\mcitedefaultseppunct}\relax
\EndOfBibitem
\bibitem[Li \emph{et~al.}(2010)Li, Chen, He, and Kang]{Li10}
Z.~Li, Z.-X. Chen, X.~He and G.-J. Kang, \emph{J. Chem. Phys.}, 2010,
  \textbf{132}, 184702/1--5\relax
\mciteBstWouldAddEndPuncttrue
\mciteSetBstMidEndSepPunct{\mcitedefaultmidpunct}
{\mcitedefaultendpunct}{\mcitedefaultseppunct}\relax
\EndOfBibitem
\bibitem[Yang \emph{et~al.}(2010)Yang, Wang, Wang, Zhang, and Li]{Yang10}
X.-F. Yang, A.-Q. Wang, Y.-L. Wang, T.~Zhang and J.~Li, \emph{J. Phys. Chem.
  C}, 2010, \textbf{114}, 3131--3139\relax
\mciteBstWouldAddEndPuncttrue
\mciteSetBstMidEndSepPunct{\mcitedefaultmidpunct}
{\mcitedefaultendpunct}{\mcitedefaultseppunct}\relax
\EndOfBibitem
\bibitem[Boronat \emph{et~al.}(2009)Boronat, Illas, and Corma]{Boronat09}
M.~Boronat, F.~Illas and A.~Corma, \emph{J. Phys. Chem. A}, 2009, \textbf{113},
  3750--3757\relax
\mciteBstWouldAddEndPuncttrue
\mciteSetBstMidEndSepPunct{\mcitedefaultmidpunct}
{\mcitedefaultendpunct}{\mcitedefaultseppunct}\relax
\EndOfBibitem
\bibitem[Florez \emph{et~al.}(2010)Florez, Gomez, Liu, Rodr\'{i}guez, and
  Illas]{Florez10}
E.~Florez, T.~Gomez, P.~Liu, J.~A. Rodr\'{i}guez and F.~Illas, \emph{Chem. Cat.
  Chem.}, 2010, \textbf{2}, 1219--1222\relax
\mciteBstWouldAddEndPuncttrue
\mciteSetBstMidEndSepPunct{\mcitedefaultmidpunct}
{\mcitedefaultendpunct}{\mcitedefaultseppunct}\relax
\EndOfBibitem
\bibitem[Perdew \emph{et~al.}(1996)Perdew, Burke, and Ernzerhof]{Perdew96}
J.~P. Perdew, K.~Burke and M.~Ernzerhof, \emph{Phys. Rev. Lett.}, 1996,
  \textbf{77}, 3865--3868\relax
\mciteBstWouldAddEndPuncttrue
\mciteSetBstMidEndSepPunct{\mcitedefaultmidpunct}
{\mcitedefaultendpunct}{\mcitedefaultseppunct}\relax
\EndOfBibitem
\bibitem[Artacho \emph{et~al.}(1999)Artacho, S\'{a}nchez-Portal, Ordej\'{o}n,
  Garc\'{i}a, and Soler]{Artacho99}
E.~Artacho, D.~S\'{a}nchez-Portal, P.~Ordej\'{o}n, A.~Garc\'{i}a and J.~M.
  Soler, \emph{Phys. Stat. Sol. (b)}, 1999, \textbf{215}, 809--817\relax
\mciteBstWouldAddEndPuncttrue
\mciteSetBstMidEndSepPunct{\mcitedefaultmidpunct}
{\mcitedefaultendpunct}{\mcitedefaultseppunct}\relax
\EndOfBibitem
\bibitem[Junquera \emph{et~al.}(2001)Junquera, Paz, S\'{a}nchez-Portal, and
  Artacho]{Junquera01}
J.~Junquera, O.~Paz, D.~S\'{a}nchez-Portal and E.~Artacho, \emph{Phys. Rev. B},
  2001, \textbf{64}, 235111/1--9\relax
\mciteBstWouldAddEndPuncttrue
\mciteSetBstMidEndSepPunct{\mcitedefaultmidpunct}
{\mcitedefaultendpunct}{\mcitedefaultseppunct}\relax
\EndOfBibitem
\bibitem[Troullier and Martins(1991)]{Troullier91}
N.~Troullier and J.~L. Martins, \emph{Phys. Rev. B}, 1991, \textbf{43},
  1993--2006\relax
\mciteBstWouldAddEndPuncttrue
\mciteSetBstMidEndSepPunct{\mcitedefaultmidpunct}
{\mcitedefaultendpunct}{\mcitedefaultseppunct}\relax
\EndOfBibitem
\bibitem[Kleinman and Bylander(1982)]{Kleinman82}
L.~Kleinman and D.~M. Bylander, \emph{Phys. Rev. Lett.}, 1982, \textbf{48},
  1425--1428\relax
\mciteBstWouldAddEndPuncttrue
\mciteSetBstMidEndSepPunct{\mcitedefaultmidpunct}
{\mcitedefaultendpunct}{\mcitedefaultseppunct}\relax
\EndOfBibitem
\bibitem[S\'{a}nchez-Portal \emph{et~al.}(1997)S\'{a}nchez-Portal, Ordej\'{o}n,
  Artacho, and Soler]{Sanchez-Portal97}
D.~S\'{a}nchez-Portal, P.~Ordej\'{o}n, E.~Artacho and J.~M. Soler, \emph{Int.
  J. Quantum Chem.}, 1997, \textbf{65}, 453--461\relax
\mciteBstWouldAddEndPuncttrue
\mciteSetBstMidEndSepPunct{\mcitedefaultmidpunct}
{\mcitedefaultendpunct}{\mcitedefaultseppunct}\relax
\EndOfBibitem
\bibitem[Soler \emph{et~al.}(2002)Soler, Artacho, Gale, Garc\'{i}a, Junquera,
  Ordej\'{o}n, and S\'{a}nchez-Portal]{Soler02}
J.~M. Soler, E.~Artacho, J.~D. Gale, A.~Garc\'{i}a, J.~Junquera, P.~Ordej\'{o}n
  and D.~S\'{a}nchez-Portal, \emph{J. Phys.: Condens. Matter}, 2002,
  \textbf{14}, 2745--2779\relax
\mciteBstWouldAddEndPuncttrue
\mciteSetBstMidEndSepPunct{\mcitedefaultmidpunct}
{\mcitedefaultendpunct}{\mcitedefaultseppunct}\relax
\EndOfBibitem
\bibitem[S\'{a}nchez-Portal \emph{et~al.}(2004)S\'{a}nchez-Portal, Ordej\'{o}n,
  and Canadell]{Sanchez-Portal04}
D.~S\'{a}nchez-Portal, P.~Ordej\'{o}n and E.~Canadell, \emph{Structure and
  Bonding}, 2004, \textbf{113}, 103--170\relax
\mciteBstWouldAddEndPuncttrue
\mciteSetBstMidEndSepPunct{\mcitedefaultmidpunct}
{\mcitedefaultendpunct}{\mcitedefaultseppunct}\relax
\EndOfBibitem
\bibitem[Huber and Herzberg(1979)]{Herzberg79}
K.~P. Huber and G.~Herzberg, \emph{Molecular Spectra and Molecular Structure
  Constants of Diatomic Molecules}, Van Nostrand Reinhold, New York, 1979\relax
\mciteBstWouldAddEndPuncttrue
\mciteSetBstMidEndSepPunct{\mcitedefaultmidpunct}
{\mcitedefaultendpunct}{\mcitedefaultseppunct}\relax
\EndOfBibitem
\bibitem[Morosin(1968)]{Morosin68}
B.~Morosin, \emph{J. Appl. Cryst.}, 1968, \textbf{1}, 123--124\relax
\mciteBstWouldAddEndPuncttrue
\mciteSetBstMidEndSepPunct{\mcitedefaultmidpunct}
{\mcitedefaultendpunct}{\mcitedefaultseppunct}\relax
\EndOfBibitem
\bibitem[Haas \emph{et~al.}(2009)Haas, Tran, and Blaha]{Haas09}
P.~Haas, F.~Tran and P.~Blaha, \emph{Phys. Rev. B}, 2009, \textbf{79},
  085104/1--10\relax
\mciteBstWouldAddEndPuncttrue
\mciteSetBstMidEndSepPunct{\mcitedefaultmidpunct}
{\mcitedefaultendpunct}{\mcitedefaultseppunct}\relax
\EndOfBibitem
\bibitem[Ding \emph{et~al.}(2004)Ding, Li, Yang, Hou, and Zhu]{Ding04}
X.~Ding, Z.~Li, J.~Yang, J.~G. Hou and Q.~Zhu, \emph{J. Chem. Phys.}, 2004,
  \textbf{120}, 9594--9600\relax
\mciteBstWouldAddEndPuncttrue
\mciteSetBstMidEndSepPunct{\mcitedefaultmidpunct}
{\mcitedefaultendpunct}{\mcitedefaultseppunct}\relax
\EndOfBibitem
\bibitem[Fern\'{a}ndez \emph{et~al.}(2004)Fern\'{a}ndez, Soler, Garz\'{o}n, and
  Balb\'{a}s]{Fernandez04}
E.~Fern\'{a}ndez, J.~M. Soler, I.~L. Garz\'{o}n and L.~C. Balb\'{a}s,
  \emph{Phys. Rev. B}, 2004, \textbf{70}, 165403/1--14\relax
\mciteBstWouldAddEndPuncttrue
\mciteSetBstMidEndSepPunct{\mcitedefaultmidpunct}
{\mcitedefaultendpunct}{\mcitedefaultseppunct}\relax
\EndOfBibitem
\bibitem[Walker(2005)]{Walker05}
A.~W. Walker, \emph{J. Chem. Phys.}, 2005, \textbf{122}, 094310/1--12\relax
\mciteBstWouldAddEndPuncttrue
\mciteSetBstMidEndSepPunct{\mcitedefaultmidpunct}
{\mcitedefaultendpunct}{\mcitedefaultseppunct}\relax
\EndOfBibitem
\bibitem[Xiao \emph{et~al.}(2006)Xiao, Tollberg, Hu, and Wang]{Xiao06}
L.~Xiao, B.~Tollberg, X.~Hu and L.~Wang, \emph{J. Chem. Phys.}, 2006,
  \textbf{124}, 114309/1--10\relax
\mciteBstWouldAddEndPuncttrue
\mciteSetBstMidEndSepPunct{\mcitedefaultmidpunct}
{\mcitedefaultendpunct}{\mcitedefaultseppunct}\relax
\EndOfBibitem
\bibitem[H\"{a}kkinen(2008)]{Hakkinen08}
H.~H\"{a}kkinen, \emph{Chem. Soc. Rev.}, 2008, \textbf{37}, 1847--1859\relax
\mciteBstWouldAddEndPuncttrue
\mciteSetBstMidEndSepPunct{\mcitedefaultmidpunct}
{\mcitedefaultendpunct}{\mcitedefaultseppunct}\relax
\EndOfBibitem
\bibitem[Nelder and Mead(1965)]{Nelder65}
J.~A. Nelder and R.~Mead, \emph{The Computer Journal}, 1965, \textbf{7},
  308--313\relax
\mciteBstWouldAddEndPuncttrue
\mciteSetBstMidEndSepPunct{\mcitedefaultmidpunct}
{\mcitedefaultendpunct}{\mcitedefaultseppunct}\relax
\EndOfBibitem
\bibitem[Burdett \emph{et~al.}(1987)Burdett, Hughbanks, Miller, Richardson, and
  Smith]{Burdett87}
J.~K. Burdett, T.~Hughbanks, G.~J. Miller, J.~W. Richardson, Jr. and J.~V.
  Smith, \emph{J. Am. Chem. Soc.}, 1987, \textbf{109}, 3639--3646\relax
\mciteBstWouldAddEndPuncttrue
\mciteSetBstMidEndSepPunct{\mcitedefaultmidpunct}
{\mcitedefaultendpunct}{\mcitedefaultseppunct}\relax
\EndOfBibitem
\bibitem[Muscat \emph{et~al.}(2002)Muscat, Swamy, and Harrison]{Muscat02}
J.~Muscat, V.~Swamy and N.~M. Harrison, \emph{Phys. Rev. B}, 2002, \textbf{65},
  224112/1--15\relax
\mciteBstWouldAddEndPuncttrue
\mciteSetBstMidEndSepPunct{\mcitedefaultmidpunct}
{\mcitedefaultendpunct}{\mcitedefaultseppunct}\relax
\EndOfBibitem
\bibitem[Monkhorst and Pack(1976)]{Monkhorst-Pack}
H.~J. Monkhorst and J.~D. Pack, \emph{Phys. Rev. B}, 1976, \textbf{13},
  5188\relax
\mciteBstWouldAddEndPuncttrue
\mciteSetBstMidEndSepPunct{\mcitedefaultmidpunct}
{\mcitedefaultendpunct}{\mcitedefaultseppunct}\relax
\EndOfBibitem
\bibitem[Perron \emph{et~al.}(2007)Perron, Domain, Roques, Drot, Simoni, and
  Catalette]{Perron07}
H.~Perron, C.~Domain, J.~Roques, R.~Drot, E.~Simoni and H.~Catalette,
  \emph{Theor. Chem. Acc.}, 2007, \textbf{117}, 565--574\relax
\mciteBstWouldAddEndPuncttrue
\mciteSetBstMidEndSepPunct{\mcitedefaultmidpunct}
{\mcitedefaultendpunct}{\mcitedefaultseppunct}\relax
\EndOfBibitem
\bibitem[Bredow \emph{et~al.}(2004)Bredow, Giordano, Cinquini, and
  Pacchioni]{Bredow04}
T.~Bredow, L.~Giordano, F.~Cinquini and G.~Pacchioni, \emph{Phys. Rev. B},
  2004, \textbf{70}, 035419/1--6\relax
\mciteBstWouldAddEndPuncttrue
\mciteSetBstMidEndSepPunct{\mcitedefaultmidpunct}
{\mcitedefaultendpunct}{\mcitedefaultseppunct}\relax
\EndOfBibitem
\bibitem[Ramamoorthy \emph{et~al.}(1994)Ramamoorthy, Vanderbilt, and
  R.D.]{Ramamoorthy94}
M.~Ramamoorthy, D.~Vanderbilt and K.-S. R.D., \emph{Phys. Rev. B}, 1994,
  \textbf{49}, 16721--16727\relax
\mciteBstWouldAddEndPuncttrue
\mciteSetBstMidEndSepPunct{\mcitedefaultmidpunct}
{\mcitedefaultendpunct}{\mcitedefaultseppunct}\relax
\EndOfBibitem
\bibitem[Liu \emph{et~al.}(2006)Liu, McAllister, Ye, and Hu]{Liu06}
L.~M. Liu, B.~McAllister, H.~Q. Ye and P.~Hu, \emph{J. Am. Chem. Soc.}, 2006,
  \textbf{128}, 4017--4022\relax
\mciteBstWouldAddEndPuncttrue
\mciteSetBstMidEndSepPunct{\mcitedefaultmidpunct}
{\mcitedefaultendpunct}{\mcitedefaultseppunct}\relax
\EndOfBibitem
\bibitem[Wang and Gong(2006)]{Wang06}
Y.~Wang and X.~G. Gong, \emph{J. Chem. Phys.}, 2006, \textbf{125},
  124703/1--12\relax
\mciteBstWouldAddEndPuncttrue
\mciteSetBstMidEndSepPunct{\mcitedefaultmidpunct}
{\mcitedefaultendpunct}{\mcitedefaultseppunct}\relax
\EndOfBibitem
\bibitem[Kang \emph{et~al.}(2009)Kang, Chen, Li, and He]{Kang09a}
G.-J. Kang, Z.-X. Chen, Z.~Li and X.~He, \emph{J. Chem. Phys.}, 2009,
  \textbf{130}, 034701/1--6\relax
\mciteBstWouldAddEndPuncttrue
\mciteSetBstMidEndSepPunct{\mcitedefaultmidpunct}
{\mcitedefaultendpunct}{\mcitedefaultseppunct}\relax
\EndOfBibitem
\bibitem[Zanchet \emph{et~al.}(2009)Zanchet, Dorta-Urra, Roncero, Flores,
  Tablero, Paniagua, and Aguado]{Zanchet09}
A.~Zanchet, A.~Dorta-Urra, O.~Roncero, F.~Flores, C.~Tablero, M.~Paniagua and
  A.~Aguado, \emph{Phys. Chem. Chem. Phys.}, 2009, \textbf{11},
  10122--10131\relax
\mciteBstWouldAddEndPuncttrue
\mciteSetBstMidEndSepPunct{\mcitedefaultmidpunct}
{\mcitedefaultendpunct}{\mcitedefaultseppunct}\relax
\EndOfBibitem
\bibitem[Pyykk\"{o}(2008)]{Pyykko08}
P.~Pyykk\"{o}, \emph{Chem. Soc. Rev.}, 2008, \textbf{37}, 1967--1997\relax
\mciteBstWouldAddEndPuncttrue
\mciteSetBstMidEndSepPunct{\mcitedefaultmidpunct}
{\mcitedefaultendpunct}{\mcitedefaultseppunct}\relax
\EndOfBibitem
\bibitem[H\"{a}kkinen \emph{et~al.}(2003)H\"{a}kkinen, Yoon, Landman, Li, Zhai,
  and Wang]{Hakkinen03}
H.~H\"{a}kkinen, B.~Yoon, U.~Landman, X.~Li, H.~Zhai and L.~Wang, \emph{Phys.
  Chem. A}, 2003, \textbf{107}, 6168--6175\relax
\mciteBstWouldAddEndPuncttrue
\mciteSetBstMidEndSepPunct{\mcitedefaultmidpunct}
{\mcitedefaultendpunct}{\mcitedefaultseppunct}\relax
\EndOfBibitem
\bibitem[Assadollahzadeh and Schwerdtfeger(2009)]{Ass09}
B.~Assadollahzadeh and P.~Schwerdtfeger, \emph{J. Chem. Phys.}, 2009,
  \textbf{131}, 064306/1--11\relax
\mciteBstWouldAddEndPuncttrue
\mciteSetBstMidEndSepPunct{\mcitedefaultmidpunct}
{\mcitedefaultendpunct}{\mcitedefaultseppunct}\relax
\EndOfBibitem
\bibitem[H\"{a}kkinen \emph{et~al.}(2002)H\"{a}kkinen, Moseler, and
  Landman]{Hakkinen02}
H.~H\"{a}kkinen, M.~Moseler and U.~Landman, \emph{Phys. Rev. Lett.}, 2002,
  \textbf{89}, 033401/1--4\relax
\mciteBstWouldAddEndPuncttrue
\mciteSetBstMidEndSepPunct{\mcitedefaultmidpunct}
{\mcitedefaultendpunct}{\mcitedefaultseppunct}\relax
\EndOfBibitem
\bibitem[Olson \emph{et~al.}(2005)Olson, Varganov, Gordon, Metiu, Chretien,
  Piecuch, Kowalski, Kucharski, and Musial]{Olson05}
R.~M. Olson, S.~Varganov, M.~S. Gordon, H.~Metiu, S.~Chretien, P.~Piecuch,
  K.~Kowalski, S.~A. Kucharski and M.~Musial, \emph{J. Am. Chem. Soc.}, 2005,
  \textbf{127}, 1049--1052\relax
\mciteBstWouldAddEndPuncttrue
\mciteSetBstMidEndSepPunct{\mcitedefaultmidpunct}
{\mcitedefaultendpunct}{\mcitedefaultseppunct}\relax
\EndOfBibitem
\bibitem[Han(2006)]{Han06}
Y.-K. Han, \emph{J. Chem. Phys.}, 2006, \textbf{124}, 024316/1--3\relax
\mciteBstWouldAddEndPuncttrue
\mciteSetBstMidEndSepPunct{\mcitedefaultmidpunct}
{\mcitedefaultendpunct}{\mcitedefaultseppunct}\relax
\EndOfBibitem
\bibitem[Choi \emph{et~al.}(2009)Choi, Kim, Lee, and Kim]{Choi09}
Y.~C. Choi, W.~Y. Kim, H.~M. Lee and K.~S. Kim, \emph{J. Chem. Theory Comput.},
  2009, \textbf{5}, 1216--1223\relax
\mciteBstWouldAddEndPuncttrue
\mciteSetBstMidEndSepPunct{\mcitedefaultmidpunct}
{\mcitedefaultendpunct}{\mcitedefaultseppunct}\relax
\EndOfBibitem
\bibitem[Mart\'{i}nez(2010)]{Martinez10}
A.~Mart\'{i}nez, \emph{J. Phys. Chem. C}, 2010, \textbf{114},
  21240--21246\relax
\mciteBstWouldAddEndPuncttrue
\mciteSetBstMidEndSepPunct{\mcitedefaultmidpunct}
{\mcitedefaultendpunct}{\mcitedefaultseppunct}\relax
\EndOfBibitem
\bibitem[Lyalin and Taketsugu(2009)]{Lyalin09a}
A.~Lyalin and T.~Taketsugu, \emph{AIP Conference Proceedings}, 2009,
  \textbf{1197}, 65--75\relax
\mciteBstWouldAddEndPuncttrue
\mciteSetBstMidEndSepPunct{\mcitedefaultmidpunct}
{\mcitedefaultendpunct}{\mcitedefaultseppunct}\relax
\EndOfBibitem
\bibitem[Semenikhina \emph{et~al.}(2008)Semenikhina, Lyalin, Solov'yov, and
  Greiner]{Semenikhina08}
V.~V. Semenikhina, A.~Lyalin, A.~V. Solov'yov and W.~Greiner, \emph{JETP},
  2008, \textbf{106}, 678--689\relax
\mciteBstWouldAddEndPuncttrue
\mciteSetBstMidEndSepPunct{\mcitedefaultmidpunct}
{\mcitedefaultendpunct}{\mcitedefaultseppunct}\relax
\EndOfBibitem
\bibitem[Ricci \emph{et~al.}(2006)Ricci, Bongiorno, Pacchioni, and
  Landman]{Ricci06}
D.~Ricci, A.~Bongiorno, G.~Pacchioni and U.~Landman, \emph{Phys. Rev. Lett.},
  2006, \textbf{97}, 036106/1--4\relax
\mciteBstWouldAddEndPuncttrue
\mciteSetBstMidEndSepPunct{\mcitedefaultmidpunct}
{\mcitedefaultendpunct}{\mcitedefaultseppunct}\relax
\EndOfBibitem
\bibitem[Bader(1990)]{Bader90}
R.~Bader, \emph{Atoms in Molecules: A Quantum Theory}, Oxford University Press,
  New York, 1990\relax
\mciteBstWouldAddEndPuncttrue
\mciteSetBstMidEndSepPunct{\mcitedefaultmidpunct}
{\mcitedefaultendpunct}{\mcitedefaultseppunct}\relax
\EndOfBibitem
\bibitem[Henkelman \emph{et~al.}(2006)Henkelman, Arnaldsson, and
  J\`{o}nsson]{Henkelman06}
G.~Henkelman, A.~Arnaldsson and H.~J\`{o}nsson, \emph{Comput. Mater. Sci.},
  2006, \textbf{36}, 354--360\relax
\mciteBstWouldAddEndPuncttrue
\mciteSetBstMidEndSepPunct{\mcitedefaultmidpunct}
{\mcitedefaultendpunct}{\mcitedefaultseppunct}\relax
\EndOfBibitem
\bibitem[Br\'{e}chignac \emph{et~al.}(1996)Br\'{e}chignac, Cahuzac, de~Frutos,
  K\'{e}ba\"{i}li, Sarfati, and Akulin]{Brechignac_PRL96}
C.~Br\'{e}chignac, P.~Cahuzac, M.~de~Frutos, N.~K\'{e}ba\"{i}li, A.~Sarfati and
  V.~Akulin, \emph{Phys. Rev. Lett.}, 1996, \textbf{77}, 251–254\relax
\mciteBstWouldAddEndPuncttrue
\mciteSetBstMidEndSepPunct{\mcitedefaultmidpunct}
{\mcitedefaultendpunct}{\mcitedefaultseppunct}\relax
\EndOfBibitem
\bibitem[Br\'{e}chignac \emph{et~al.}(1998)Br\'{e}chignac, Cahuzac,
  K\'{e}ba\"{i}li, and Leygnier]{Brechignac_PRL98}
C.~Br\'{e}chignac, P.~Cahuzac, N.~K\'{e}ba\"{i}li and J.~Leygnier, \emph{Phys.
  Rev. Lett.}, 1998, \textbf{81}, 4612--4615\relax
\mciteBstWouldAddEndPuncttrue
\mciteSetBstMidEndSepPunct{\mcitedefaultmidpunct}
{\mcitedefaultendpunct}{\mcitedefaultseppunct}\relax
\EndOfBibitem
\bibitem[Obolensky \emph{et~al.}(2005)Obolensky, Lyalin, Solov'yov, and
  Greiner]{Obolensky05}
O.~I. Obolensky, A.~G. Lyalin, A.~V. Solov'yov and W.~Greiner, \emph{Phys. Rev.
  B}, 2005, \textbf{72}, 085433/1--11\relax
\mciteBstWouldAddEndPuncttrue
\mciteSetBstMidEndSepPunct{\mcitedefaultmidpunct}
{\mcitedefaultendpunct}{\mcitedefaultseppunct}\relax
\EndOfBibitem
\bibitem[Yoon \emph{et~al.}(2005)Yoon, H\"{a}kkinen, Landman, W\"{o}rz,
  Antonietti, Abbet, Judai, and Heiz]{Yoon05}
B.~Yoon, H.~H\"{a}kkinen, U.~Landman, A.~S. W\"{o}rz, J.-M. Antonietti,
  S.~Abbet, K.~Judai and U.~Heiz, \emph{Science}, 2005, \textbf{307},
  403--407\relax
\mciteBstWouldAddEndPuncttrue
\mciteSetBstMidEndSepPunct{\mcitedefaultmidpunct}
{\mcitedefaultendpunct}{\mcitedefaultseppunct}\relax
\EndOfBibitem
\bibitem[Chr\'{e}tien and Metiu(2007)]{Chretien07}
S.~Chr\'{e}tien and H.~Metiu, \emph{J. Chem. Phys.}, 2007, \textbf{126},
  104701/1--7\relax
\mciteBstWouldAddEndPuncttrue
\mciteSetBstMidEndSepPunct{\mcitedefaultmidpunct}
{\mcitedefaultendpunct}{\mcitedefaultseppunct}\relax
\EndOfBibitem
\end{mcitethebibliography}
\bibliographystyle{rsc}
}

\end{document}